\documentclass[twocolumn,cleanfoot,10pt]{asme2ej}

\usepackage{graphics} % for pdf, bitmapped graphics files
\usepackage{epsfig} % for postscript graphics files
\usepackage{mathptmx} % assumes new font selection scheme installed
\usepackage{times} % assumes new font selection scheme installed

\usepackage{amsmath,amssymb,amsthm}
\usepackage{caption}
\usepackage[figbotcap]{subfigure}
\usepackage{color}
\usepackage{gensymb}%for the \degree command
\usepackage{siunitx}
\usepackage{multirow}
\usepackage{bm}
\usepackage{ifthen}
\usepackage{url}

\theoremstyle{remark} %3 possibilities: plain (italic), definition,
\newtheorem{comment}{Comment}

\def\rd#1{{\color{red}{#1}}}
\def\pb#1{\footnote{PB:\rd{#1}}}

\def\eqdef{\ensuremath{:=}}
\def\defeq{\ensuremath{=:}}

\def\bl{$BL$}
\def\slmpc{$MZHC$}
\def\shyphen{$-$}
\def\hlc{HLC}
\def\llc{LLC}
\def\hiresmodel{high-resolution}
\def\Hiresmodel{High-resolution}
\def\loresmodel{low-resolution}
\def\metazone{meta-zone}
\def\metazones{meta-zones}

\def\version{Arxiv}% Jrnl or Arxiv

\newboolean{ArxivVersion}
\ifthenelse{\equal{\version}{Arxiv}}
{\setboolean{ArxivVersion}{true} 
}
{\setboolean{ArxivVersion}{false} 
}

\usepackage{xparse}

\ExplSyntaxOn
\cs_new:Nn \pb_prop_gset_bykeys:Nn
{
	\prop_clear:N \l__pb_temp_prop
	\keys_set:nn { pb/propbykey } { #2 }
	\prop_gset_eq:NN #1 \l__pb_temp_prop
}
\cs_generate_variant:Nn \pb_prop_gset_bykeys:Nn { c }

\keys_define:nn { pb/propbykey }
{
	unknown .code:n = \prop_put:Nxn \l__pb_temp_prop { \l_keys_key_tl } { #1 }
}

\cs_generate_variant:Nn \prop_put:Nnn { Nx }

\NewDocumentCommand{\setupcollaborator}{mm}
{% #1 = identifier string, #2 = set of key-value pairs
	\prop_new:c { g_collaborator_#1_prop }
	\pb_prop_gset_bykeys:cn { g_collaborator_#1_prop } { #2 }
}

\NewDocumentCommand{\selectcollaborator}{m}
{
	\prop_map_inline:cn { g_collaborator_#1_prop }
	{
		\tl_set:cn { ##1 } { ##2 }
	}
}
\ExplSyntaxOff

\setupcollaborator{PBmac}
{% pb, macbook air 
	NAME=Prabir Mac , laptop or mac mini,
	DiCEbibPATH=/Users/pbarooah/Dropbox/Dropbox/DiCElab_bib,
	NRbibPATH = /Users/pbarooah/Dropbox/Dropbox/NSRbib,
}
\setupcollaborator{NRlinux}
{% NRlinux is Naren using the linux machine at work
	NAME=NR at lab,
	DiCEbibPATH=../../DiCElab_bib,
	NRbibPATH = ../../NSRbib,
}
\setupcollaborator{NRwindows}
{% NRwindows is Naren using his windows.
	NAME=NR using his windows computer,
	DiCEbibPATH=../../DiCElab_bib,
	NRbibPATH = ../../NSRbib,
}
\selectcollaborator{NRwindows}
\graphicspath{{figures/}}

\title{MPC-Based Hierarchical Control of a Multi-Zone Commercial HVAC System}

\author{Naren Srivaths Raman\thanks{Corresponding author. Email: narensraman@ufl.edu}, Rahul Umashankar Chaturvedi, Zhong Guo, and Prabir Barooah
    \affiliation{Department of Mechanical and Aerospace Engineering\\
	University of Florida\\
	Gainesville, FL 32611
   }	
}

\begin{document}

\maketitle    

%%%%%%%%%%%%%%%%%%%%%%%%%%%%%%%%%%%%%%%%%%%%%%%%%%%%%%%%%%%%%%%%%%%%%%
\begin{abstract}
{\it This paper presents a novel architecture for model predictive control (MPC) based indoor climate control of multi-zone buildings to provide energy efficiency. 	Unlike prior works we do not assume the availability of a \hiresmodel{} multi-zone building model, which is challenging to obtain.	Instead, the architecture uses a \loresmodel{} model of the building which is divided into a small number of ``meta-zones'' that can be easily identified using existing data-driven modeling techniques. The proposed architecture is hierarchical. At the higher level, an MPC controller uses the \loresmodel{} model to make decisions for the air handling unit~(AHU) and the meta-zones. Since the meta-zones are fictitious, a lower level controller converts the high-level MPC decisions into commands for the individual zones by solving a projection problem that strikes a trade-off between two potentially conflicting goals: the AHU-level decisions made by the MPC are respected while the climate of the individual zones is maintained within the comfort bounds. The performance of the proposed controller is assessed via simulations in a high-fidelity simulation testbed and compared to that of a rule-based controller that is used in practice. Simulations in multiple weather conditions show the effectiveness of the proposed controller in terms of energy savings, climate control, and computational tractability. 
}
\end{abstract}

%%%%%%%%%%%%%%%%%%%%%%%%%%%%%%%%%%%%%%%%%%%%%%%%%%%%%%%%%%%%%%%%%%%%%%
%\begin{nomenclature}
%\entry{A}{You may include nomenclature here.}
%\entry{$\alpha$}{There are two arguments for each entry of the nomemclature environment, the symbol and the definition.}
%\end{nomenclature}
%
%The primary text heading is  boldface and flushed left with the left margin.  The spacing between the  text and the heading is two line spaces.

%%%%%%%%%%%%%%%%%%%%%%%%%%%%%%%%%%%%%%%%%%%%%%%%%%%%%%%%%%%%%%%%%%%%%%
\sloppy

\section{Introduction}\label{sec:intro}
The application of model predictive control (MPC) for commercial heating, ventilation, and air conditioning (HVAC) systems for both energy efficiency and demand flexibility has been an active area of research; see the review articles~\cite{serale2018model,shaikh2014review} and the references therein. 

Several of the MPC formulations proposed in the past are for buildings with one zone~\cite{goybar:CDC:2013,JoeModel:2019,RamanMPC_AE:2020,ChenOccupant:2016} or a small number of zones~\cite{ma2012demand,Bengeaetal:2013}. A direct extension of such formulations for large multi-zone buildings has two main challenges. First, solving the underlying optimization problem in MPC for a building with a large number of zones is computationally complex because of the large number of decision variables. To reduce the computational complexity, several distributed and hierarchical approaches have been proposed~\cite{RadhakrishnanToken:2016,PngInternet:2019,ma2012distributed,mei2018multizone,patel2016distributed,yang2020hvac,long2016hierarchical}. The second challenge, which has attracted far less attention, is that MPC requires a ``\hiresmodel{}'' model of the thermal dynamics of a multi-zone building. 
\Hiresmodel{} means that the temperature of every zone in the building is a state in the model and the control commands for every zone are inputs in the model. One way of obtaining such a multi-zone model is by first constructing a ``white box'' model, such as by using a building energy modeling software, and then simplifying it to make it suitable for MPC, e.g.,~\cite{jorissen2016towards}. But constructing a white box model is expensive; it requires significant effort~\cite{liwen:2014}. Moreover, the resulting model may not reflect the building as is. Another way of obtaining a \hiresmodel{} multi-zone model is by utilizing data-driven techniques, which use input-output measurements. Getting reliable estimates using data-driven modeling is challenging even for a single-zone building, as a building's thermal dynamics is affected by a non-trivial and unmeasurable disturbance, the heat gain from occupants and their use of equipment, that strongly affects quality of the identified model~\cite{Kim_Braun:2016}. In the case of multi-zone model identification, it becomes intractable since the model has too many degrees of freedom: as many unknown disturbance signals as there are number of zones. To the best of our knowledge, there are no works on reliable identification of multi-zone building models without making assumptions on the nature of the disturbance affecting individual zones~\cite{ZengIdentification_TCST:2019}.

In addition to the challenges mentioned above, most of the prior works---whether on single-zone or on multi-zone buildings---ignore humidity and latent heat in their MPC formulations. The inclusion of moisture requires a computationally convenient cooling and dehumidifying coil model. MPC formulations which exclude humidity can lead to poor humidity control, or higher energy usage as they are unaware of the latent load on the cooling coil~\cite{RamanMPC_AE:2020}. 
\begin{figure*}[t]
	\centering
	\includegraphics[width=0.69\linewidth]{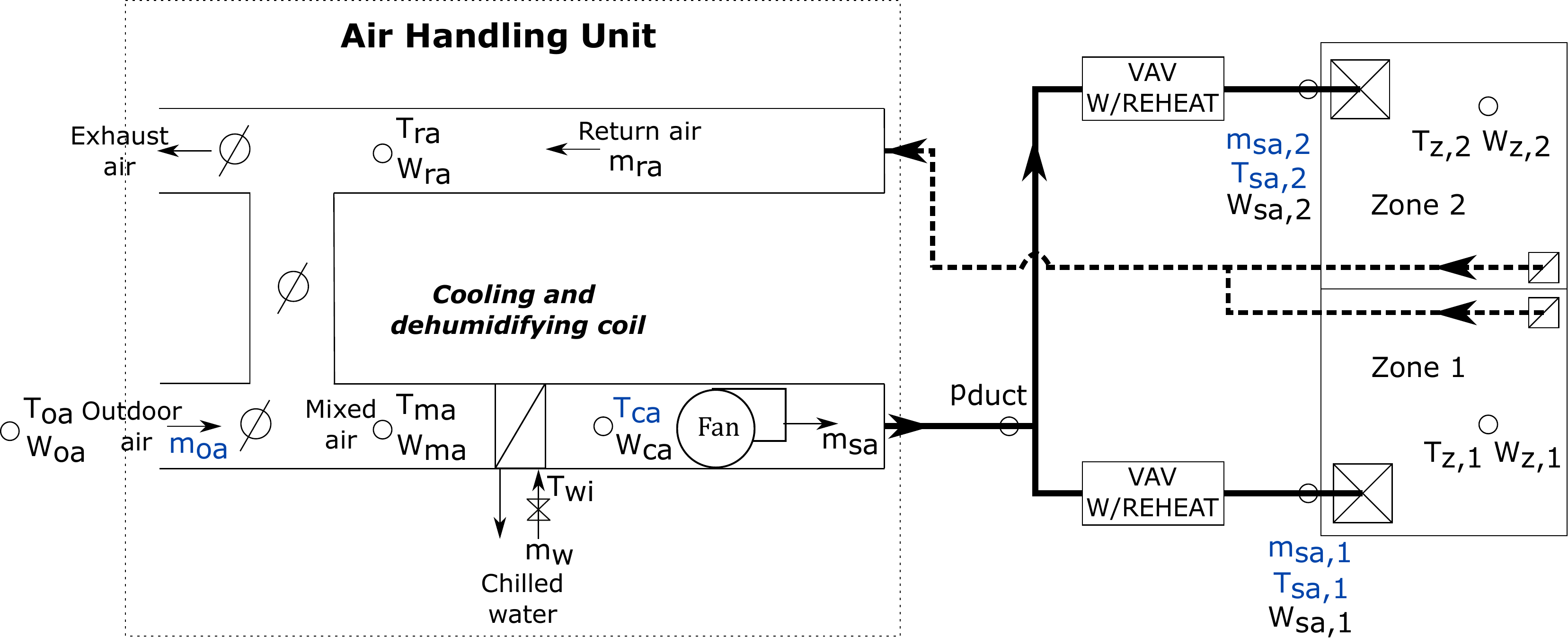}
	\caption{Schematic of a multi-zone---specifically, a two zone---commercial variable-air-volume
		HVAC system. In this figure, oa:~outdoor air, ra:~return air, ma:~mixed air, ca:~conditioned air, and sa:~supply air.}\vspace{-10pt}
	\label{fig:AHU_schematic_multi_zone}
\end{figure*}

In this work, we propose a humidity-aware MPC formulation for a multi-zone building with a variable air volume~(VAV) HVAC system. Figure~\ref{fig:AHU_schematic_multi_zone} shows the schematic of such a system.  

To overcome the challenges mentioned above, we propose a two-level control architecture. The high-level controller~(\hlc{}) decides on the AHU-level control commands. The \hlc{} is an MPC controller that uses a ``\loresmodel{}'' model of the building with a small number of ``meta-zones'', with each meta-zone being a single-zone equivalent of a part of the building consisting of several zones. In the case study presented here, a 33 zone three-floor building is aggregated to a 3 meta-zone model, with each meta-zone corresponding to a floor. The advantage of such an approach is that a \hiresmodel{} multi-zone model is not needed as a starting point. Rather, a single-zone equivalent model of each meta zone,  in which disturbance in all the zones are aggregated into one signal, can be directly identified from measurements collected from the building. The identification problem of such a single-zone equivalent model is more tractable~\cite{GuoAggregationEnB:2020}. In this paper, we use the system identification method from~\cite{GuoAggregationEnB:2020}, though other identification methods can also be used. Since the \hlc\ uses a \loresmodel{} model with a much smaller number of meta-zones than that in the building, its computational complexity is low. However, this reduction of computational complexity creates a different challenge. Since the decision variables of the optimization problem in the \hlc\ correspond to the meta-zones (air flow rate, temperature, etc.), they do not correspond to those for the actual zones of the building. The low-level controller~(\llc{}) is now used to compute the control commands for individual zones. It does so by solving a projection problem that appropriately distributes aggregate quantities computed by the \hlc{} to individual zones. The \llc{} uses feedback from each zone to assess their needs and ensures indoor climate of each zone is maintained.

The proposed controller---that includes the \hlc\ and \llc\---is hereafter referred to as \slmpc{} which stands for \emph{multi-zone hierarchical controller}. Its performance is assessed through simulations on a ``virtual building'' plant. The plant is representative of a section of the Innovation Hub building comprising of 33 zones and is located at the University of Florida campus. The plant is constructed using Modelica~\cite{fritzson1998modelica}.  The performance of the proposed controller is compared with that of \emph{Dual Maximum} controller as a baseline~\cite{ASHRAE_handbook_applications:11}. The dual maximum controller---which is referred to as \bl{} (for baseline)---is a  rule-based controller, and is one of the more energy efficient controllers among those used in practice~\cite{ASHRAE_handbook_applications:11}.  Simulation results show that using the proposed controller provides significant energy savings when compared to \bl{} while maintaining indoor climate. 

Compared to the literature on MPC design for multi-zone building HVAC systems, our work makes four principal contributions, with details discussed in Section~\ref{sec:lit}. (i)~The first contribution is that the proposed method does not assume availability of a \hiresmodel{} model of the multi-zone building which is difficult to obtain. Instead, it can utilize existing data-driven methods that can quickly identify a \loresmodel{} model of the multi-zone building from measurements. (ii)~Since the MPC part of the proposed controller uses a \loresmodel{} model with a small number of meta-zones, the method is scalable to buildings with a large number of zones.
Although distributed iterative computation has been proposed in the literature as an alternate approach to reducing computational complexity, ours can be solved in a centralized setting.
(iii)~The third contribution is the incorporation of humidity and latent heat in our multi-zone MPC formulation, which has been largely ignored in the literature on MPC for buildings, and especially so in the literature on multi-zone building MPC. Our simulations show that when using \slmpc{}, the indoor humidity constraint is active, especially during hot-humid weather. Without humidity being explicitly considered, the controller would have caused high space humidity in an effort to reduce energy use. (iv)~The fourth contribution is a realistic evaluation of the proposed controller in a high-fidelity simulation platform that introduces a large plant-model mismatch. In many prior works on multi-zone MPC, the model used by the controller is the same as that used in simulating the plant. In contrast, the only information provided to the proposed controller about the building is sensor measurements (past data for model identification and real-time data during control) and design parameters such as expected occupancy, minimum design airflow rates for each VAV box, etc.

The rest of this paper is organized as follows. Section \ref{sec:lit} discusses our work in relation to the literature on multi-zone MPC. Section~\ref{section:system_description_models} describes a multi-zone building equipped with a VAV HVAC system and the models we use in simulating the plant (the system to be controlled). Section~\ref{sec:mpc} presents the proposed MPC-based hierarchical controller. Section~\ref{sec:baseline} describes a rule-based baseline controller with which the performance of the proposed controller is compared. The simulation setup is described in Section~\ref{sec:simulation_setup}. Simulation results are presented and discussed in Section~\ref{sec:results}. Finally, the main conclusions are provided in Section~\ref{sec:conclusions}.

\subsection{Comparison With Literature on Multi-Zone MPC}\label{sec:lit}
Several distributed and hierarchical approaches have been proposed to reduce the computational complexity of MPC for multi-zone buildings~\cite{RadhakrishnanToken:2016,PngInternet:2019,ma2012distributed,mei2018multizone,patel2016distributed,yang2020hvac,long2016hierarchical}. In~\cite{RadhakrishnanToken:2016}, a hierarchical distributed algorithm called token-based scheduling is proposed to vary the supply airflow rate to the zones. A modified version of this algorithm is used in~\cite{PngInternet:2019} to minimize the energy consumption of a multi-zone building located at the Nanyang Technological University, Singapore campus.

In~\cite{mei2018multizone}, a two-layered control architecture is proposed for operating a VAV HVAC system. The upper layer is an open loop controller, while the lower layer is based on MPC and it varies the supply airflow rates to the zones. Similar to~\cite{RadhakrishnanToken:2016}, \cite{PngInternet:2019}, and~\cite{mei2018multizone}, the works~\cite{yang2020hvac,patel2016distributed,long2016hierarchical} consider varying only the zone-level control inputs such as the supply airflow rates and zone temperature set points. These works exclude the AHU-level control inputs such as the conditioned air temperature and outside airflow rate.
Unlike the works mentioned above, the work~\cite{liang2015mpc} uses MPC to vary only the AHU-level control inputs; the zone-level control inputs are excluded in this formulation. 

One of the few works similar to ours is~\cite{ma2012distributed}, as they consider both the zone-level and AHU-level control inputs in their formulation. But their algorithm requires a \hiresmodel{} multi-zone model, and they do not consider humidity and latent heat in their formulation.

\section{System and Problem Description, and Plant Simulator}\label{section:system_description_models}
Our focus is a multi-zone building equipped with a variable-air-volume (VAV) HVAC system, whose schematic is shown in Figure~\ref{fig:AHU_schematic_multi_zone}. In such a system, part of the air exhausted from the zones is recirculated and mixed with outdoor air. This mixed air is sent through the cooling coil where the air is cooled and dehumidified to the conditioned air temperature~($T_{ca}$) and humidity ratio~($W_{ca}$). This conditioned air is then sent through the supply ducts to the VAV boxes, which have a damper to control airflow, and finally supplied to the zones. Some VAV boxes have reheat coils; they can change temperature of supply air but not humidity, i.e., $T_{sa,i}\geq T_{ca}$ and $W_{sa,i}=W_{ca}$, where $T_{sa,i}$ and $W_{sa,i}$ are the temperature and humidity ratio of supply air to the $i^{th}$ zone. If a VAV box is not equipped with a reheat coil~(cooling only), then the temperature of air supplied by it to its zone will be at the conditioned air temperature, i.e., $T_{sa,i}=T_{ca}$.

The control commands for a multi-zone VAV HVAC system with $n_z$ zones (i.e., VAV boxes) are: 
\begin{align}\label{eq:u_all}
	u &\eqdef (m_{oa}, T_{ca}, m_{sa,i}, T_{sa,i},\,\,i=1,\dots,n_z),
\end{align}
where $m_{oa}$ is the outdoor airflow rate, $T_{ca}$ is the conditioned air temperature, $m_{sa,i}$ is the supply airflow rate to the $i^{th}$ zone, and $T_{sa,i}$ is the supply air temperature to the $i^{th}$ zone. Note that the humidity of conditioned air ($W_{ca}$) which is supplied to all the zones is indirectly controlled through $T_{ca}$. Of the $n_z$ VAVs/zones in the  building, $n_z^{rh}$ VAVs are equipped with a reheat coil and $n_z-n_z^{rh}$ VAVs do not have a reheat coil (cooling only). For the latter, the supply air temperature will be the same as the conditioned air temperature, i.e., $T_{sa,i}(k)=T_{ca}(k)$.

The control commands in \eqref{eq:u_all} are sent as set points to the low-level control loops which are typically comprised of proportional integral (PI) controllers. The role of a climate control system is to vary these control commands so that three main goals are satisfied: (i) ensure thermal comfort, (ii) maintain indoor air quality, and (iii) use minimum amount of energy/cost. 

\ifArxivVersion In an HVAC system as the one shown in Figure~\ref{fig:AHU_schematic_multi_zone}, the supply duct pressure setpoint, $p_{duct}$, is also usually a command that the climate control system has to decide. We assume that the supply duct static pressure setpoint~($p_{duct}$) is controlled based on ``trim and respond'' strategy~\cite{taylor2007vav}, which is commonly used in VAV systems, including in the Innovation Hub building that we use as a case study.\fi

\subsection{Virtual Building (Simulator)}\label{sec:building_thermal_model}
\begin{figure}[tb]
	\subfigure[Picture of the Innovation Hub building (view from south to north). Phase-1 is enclosed in the dashed lines.\label{fig:innovation_hub_photo}]{\includegraphics[width=0.49\textwidth]{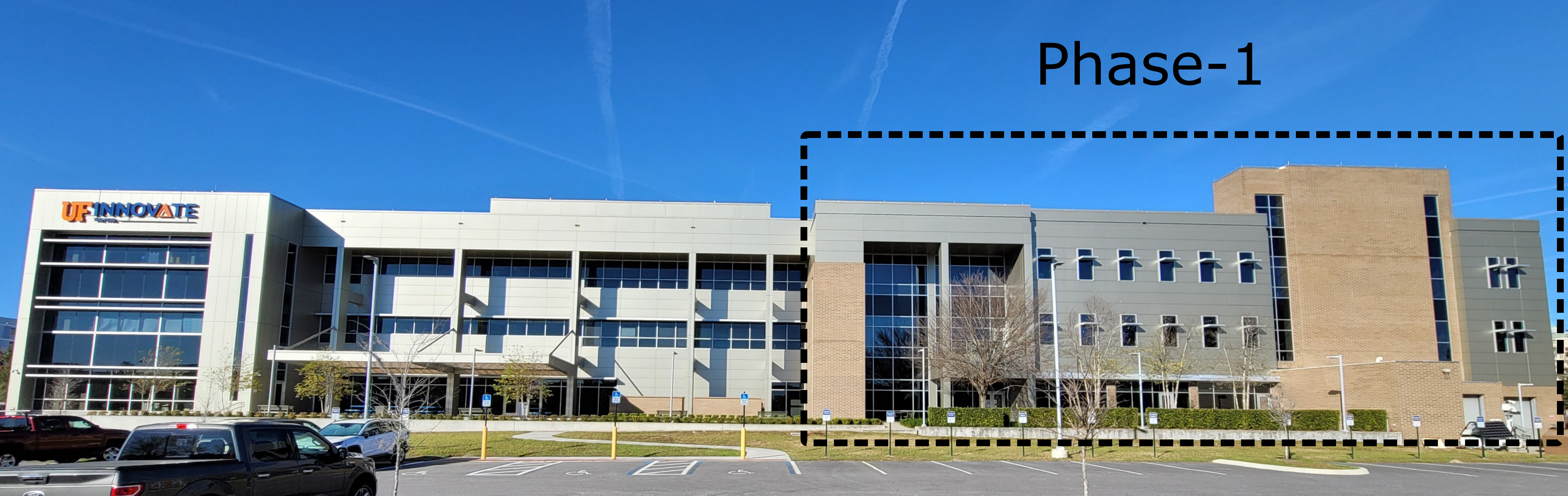}}
	\subfigure[Top view of the Innovation Hub building. In this work, we consider air handling unit~2 (AHU-2) which serves the southern half of Phase-1 (region shaded in blue). \label{fig:innovation_hub_top_view}]{\includegraphics[width=0.49\textwidth]{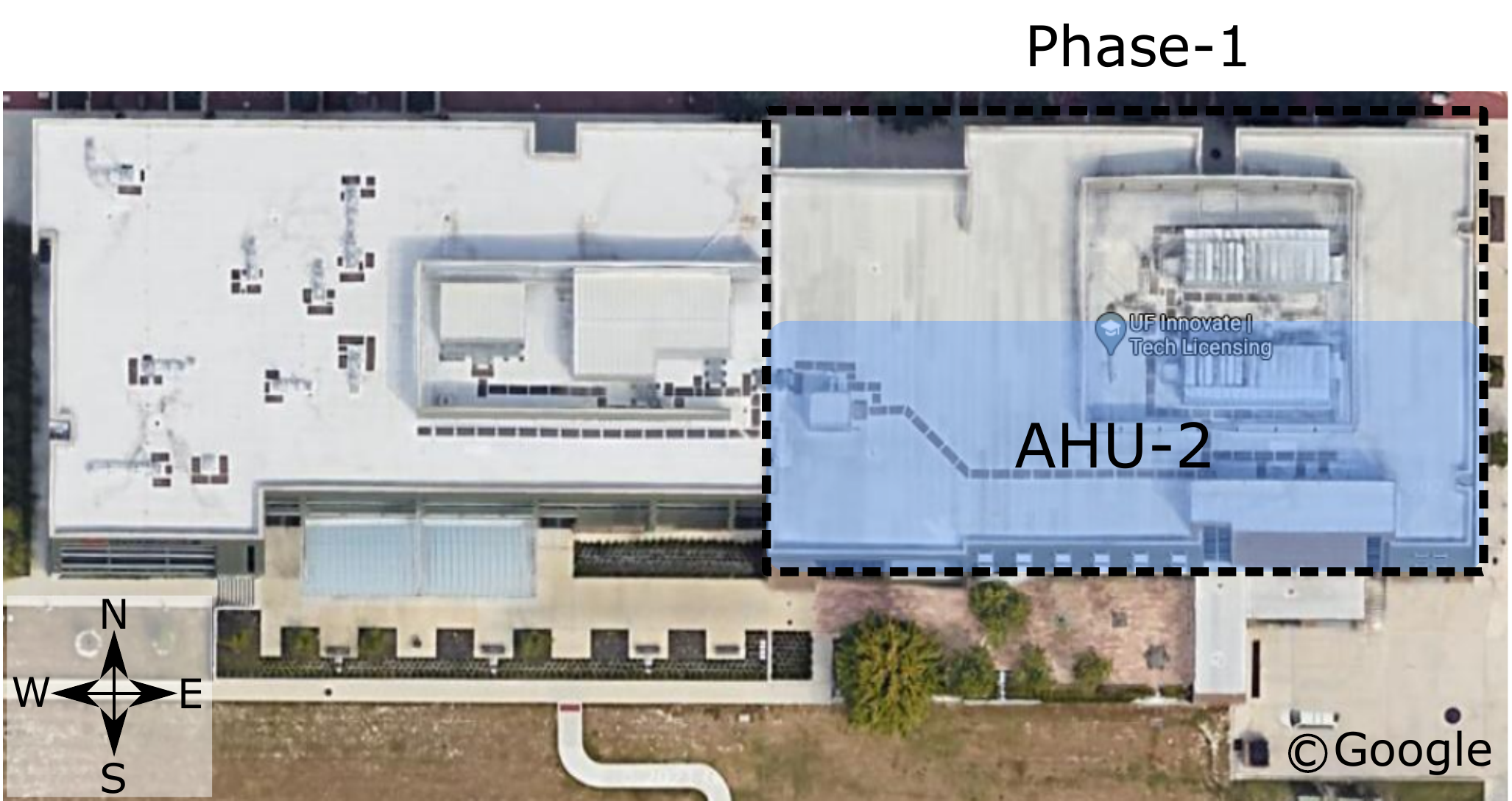}}\vspace{-10pt}
	\caption{Innovation Hub building located at the University of Florida campus.}\label{fig:innovation_hub_photos}\vspace{-10pt}
\end{figure}

\begin{figure}[tb]
	\subfigure[Floor 1 plan.\label{fig:floor_1_schematic}]{\includegraphics[width=0.49\textwidth]{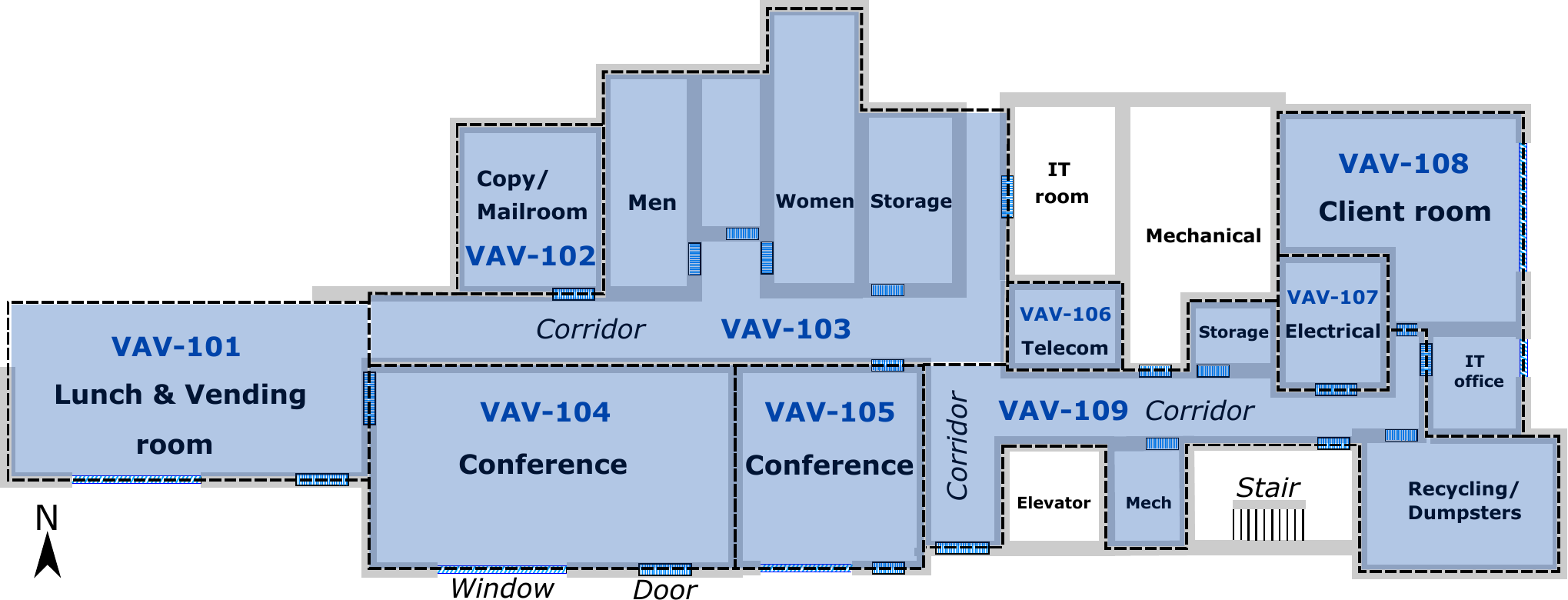}}\vspace{-5pt}
	\subfigure[Floor 2 plan.\label{fig:floor_2_schematic}]{\includegraphics[width=0.49\textwidth]{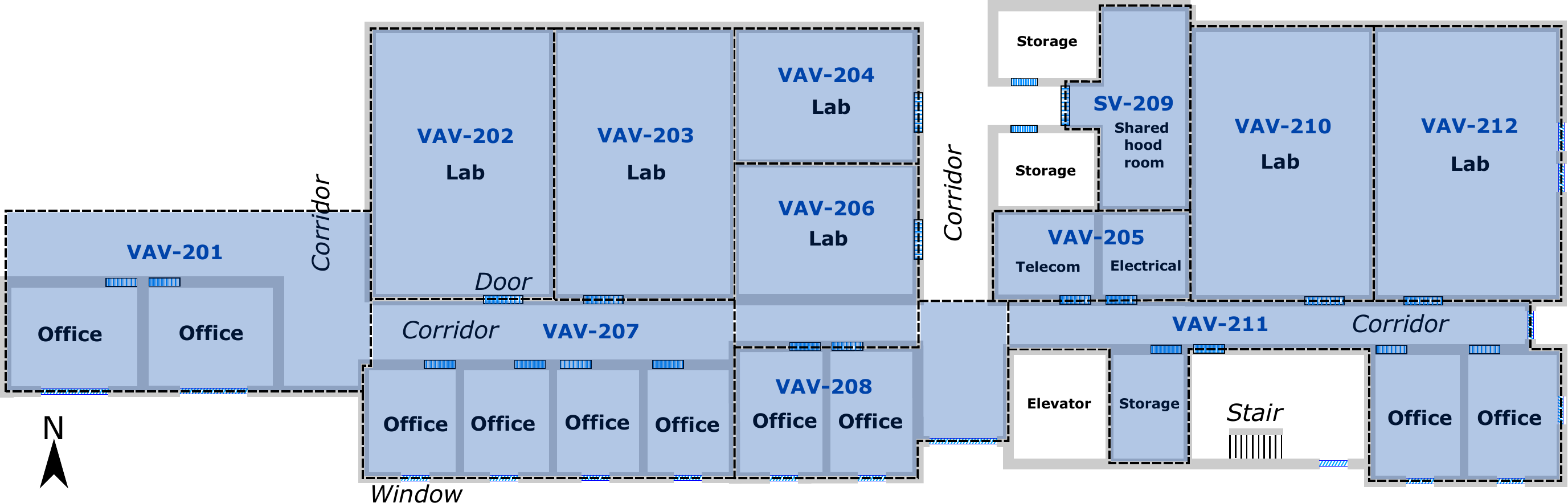}}\vspace{-5pt}\quad
	\subfigure[Floor 3 plan.\label{fig:floor_3_schematic}]{\includegraphics[width=0.49\textwidth]{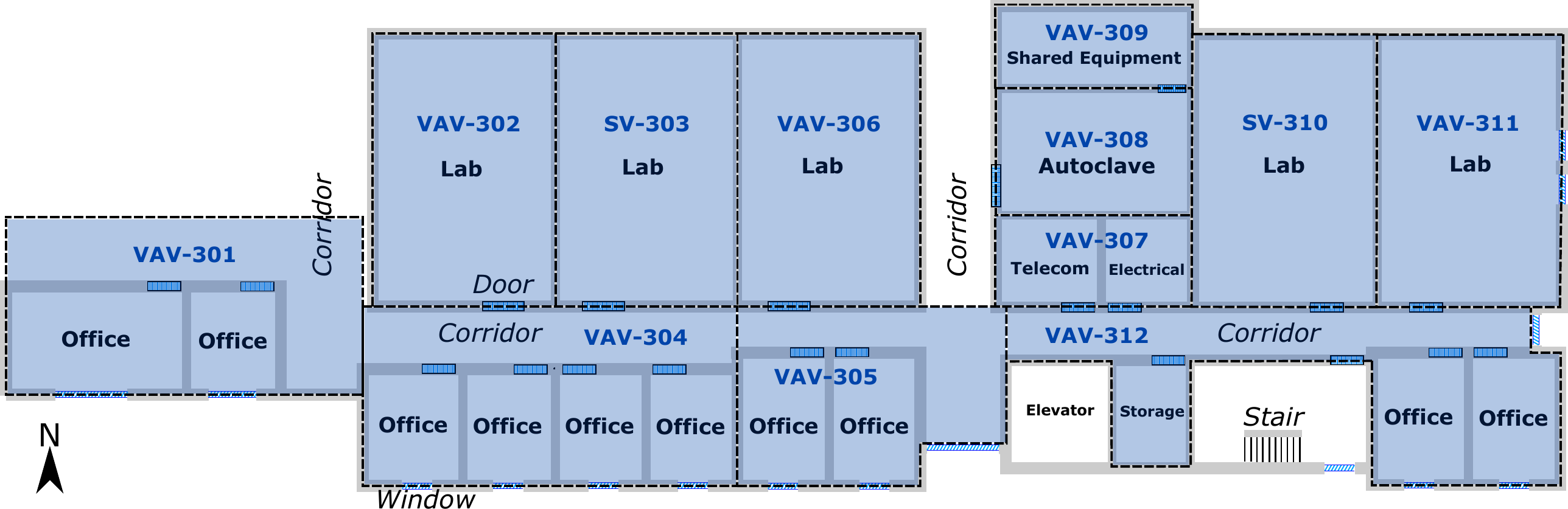}}
	\caption{Floor plans of the southern half of Phase-1 which is serviced by AHU-2.}\label{fig:floor_plans}\vspace{-10pt}
\end{figure}

The virtual building (VB) is a high-fidelity model of a building's thermal dynamics and its HVAC system that will act as the plant for the controllers. The VB is chosen to mimic part of the Innovation Hub building in Gainesville, FL, USA, which is serviced by AHU-2 (among the two AHUs that serve Phase I). Figures~\ref{fig:innovation_hub_photos} and~\ref{fig:floor_plans} show photos of the building and the relevant floor plans, respectively.  The rooms supplied by the same VAV box are grouped together to form one large space (zone); the zones are enclosed by dotted lines in Figure~\ref{fig:floor_plans}. The first floor has 15 rooms which are grouped into 9 zones, the second floor has 20 rooms which are grouped into 12 zones, and the third floor has 21 rooms which are grouped into 12 zones. In total, there are 56 rooms grouped into 33 zones. %Each of the zones are serviced by a separate VAV box.
The virtual building thus consists of an air handling unit and 33 VAV boxes, of which 29 are equipped with reheat coils, and the remaining 4 do not have reheat coils (cooling only). The zones primarily consist of offices and labs.

We use the Modelica library IDEAS (Integrated District Energy Assessment by Simulation)~\cite{jorissen2018ideas} to model the building's thermal dynamics.

\begin{figure}[t]
	\centering
	\includegraphics[width=0.99\linewidth]{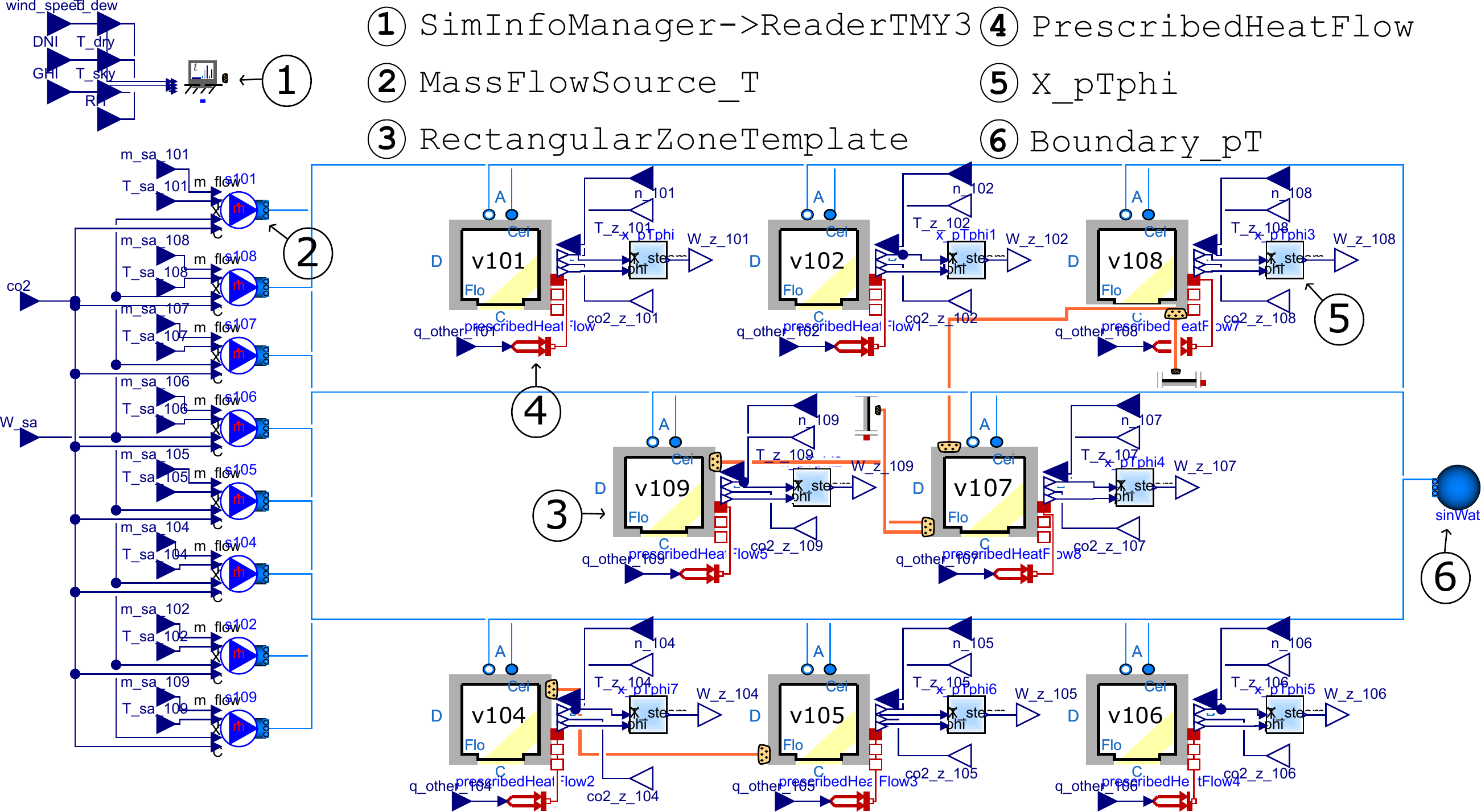}
	\caption{Floor 1 of the virtual building created in Dymola using components from the IDEAS library~\cite{jorissen2018ideas}.}
	\label{fig:dymola_floor_1_screenshot_w_legend}\vspace{-10pt}
\end{figure} 

\ifArxivVersion

In order to model a zone we use the {\tt RectangularZoneTemplate} from the IDEAS library. It consists of six components---which are a ceiling, a floor, and four walls---and an optional window. There are also external connections for each of the walls and the ceiling. Depending on the usage, there are three types of walls: (i)~inner wall, which is used as a boundary between zones, (ii)~outer wall, which is used as a boundary between outside (atmosphere) and the zone, and (iii)~boundary wall, which can be specified a fixed temperature or heat flow. To define a wall, dimensions, type of material, type of wall, and the azimuth angle are required. The dimensions are obtained from the mechanical drawings, the material type is chosen from the predefined materials available in the IDEAS library, the type of wall is chosen based on the zone's location in the building, and the azimuth angle is computed from the zone's orientation. Windows are specified according to the drawings, with the glazing material chosen from the IDEAS library. In this way, we model all the zones, which are then connected appropriately to form the overall building; Figure~\ref{fig:dymola_floor_1_screenshot_w_legend} shows the model of floor 1. Since we are only interested in modeling the southern half of Phase-1, the walls that are adjacent to zones in the northern half are assumed to be at $22.22\degree C$ ($72\degree F$).

Inputs to the building thermal dynamics portion of the VB are supply airflow rate ($m_{sa,i}$), supply air temperature ($T_{sa,i}$), and supply air humidity ratio ($W_{sa,i}$) for all the zones. These are implemented using the {\tt MassFlowSource$\_$T} block from the IDEAS library; an ideal flow source that produces a specified mass flow with specified temperature, composition, and trace substances. Outputs of the simulator are temperature ($T_{z,i}$) and humidity ratio ($W_{z,i}$) of all the zones. The zone temperature and humidity are also influenced by several exogenous inputs: (i) outdoor weather conditions such as solar irradiation ($\eta_{sol}$), outdoor air temperature ($T_{oa}$), etc. which are provided using the {\tt ReaderTMY3} block from the IDEAS library, (ii) internal sensible and latent heat loads due to occupants, which are computed based on the number of occupants provided to the zone block, and  (iii) internal heat load due to lighting and equipment which is given using the {\tt PrescribedHeatFlow} from the Modelica standard library.

\else

Inputs to the building thermal dynamics portion of the VB are supply airflow rate ($m_{sa,i}$), supply air temperature ($T_{sa,i}$), and supply air humidity ratio ($W_{sa,i}$) for all the zones, and the exogenous disturbances, which are (i) solar irradiation ($\eta_{sol}$), outdoor air temperature ($T_{oa}$), and outdoor air humidity ratio ($W_{oa}$), (ii) internal sensible and latent heat loads due to occupants, which are computed based on the number of occupants provided to the zone block, and  (iii) internal heat load due to lighting and equipment. Outputs of the simulator are temperature ($T_{z,i}$) and humidity ratio ($W_{z,i}$) of all the zones. Figure~\ref{fig:dymola_floor_1_screenshot_w_legend} shows floor~1 of the VB created in Dymola using components from the IDEAS library~\cite{jorissen2018ideas}. We omit the details of the VB construction using Modelica/Dymola; the interested reader is referred to~\cite{RamanMPC-Based_AriXiV:2021}.

\fi

\paragraph{Cooling and Dehumidifying Coil Model}
The cooling coil model has five inputs and two outputs. The inputs are supply airflow rate ($m_{sa}$), mixed air temperature ($T_{ma}$) and humidity ratio ($W_{ma}$), chilled water flow rate ($m_w$), and chilled water inlet temperature ($T_{wi}$). The outputs are conditioned air temperature ($T_{ca}$) and humidity ratio ($W_{ca}$).  
We use a gray box data-driven model developed in our prior work~\cite{RamanMPC_AE:2020}. The interested readers are referred to Section~2.1.2 of~\cite{RamanMPC_AE:2020} for details regarding the model.

\paragraph{Power Consumption Models}
For the HVAC system configuration presented in Figure~\ref{fig:AHU_schematic_multi_zone}, there are three main components which consume power. They are supply fan, cooling and dehumidifying coil, and reheating coils.
The fan power consumption is modeled as:
\begin{align}\label{eq:P_fan}
	P_{fan}(k) = \alpha_{fan} m_{sa}(k)^3,
\end{align}
where $m_{sa}(k)$ is the total supply airflow rate at the AHU~\cite{rouheiforpib:2001}.

The cooling and dehumidifying coil power consumption is modeled to be proportional to the heat it extracts from the mixed air stream:
\begin{align} \label{eq:P_cc}
	P_{cc}(k) = \frac{m_{sa}(k)\big[h_{ma}(k)-h_{ca}(k)\big]}{\eta_{cc}COP_c},
\end{align}
where $h_{ma}(k)$ and $h_{ca}(k)$ are the specific enthalpies of the mixed and conditioned air respectively, $\eta_{cc}$ is the cooling coil efficiency, and $COP_c$ is the chiller coefficient of performance. Since a part of the return air is mixed with the outside air, the specific enthalpy of the mixed air is:
\begin{align} \label{h_ma}
	h_{ma}(k) &= r_{oa}(k)h_{oa}(k) + (1-r_{oa}(k))h_{ra}(k),
\end{align}
where $h_{oa}(k)$ and $h_{ra}(k)$ are the specific enthalpies of the outdoor and return air respectively, and $r_{oa}(k)$ is the outside air ratio: $r_{oa}(k)\eqdef \frac{m_{oa}(k)}{m_{sa}(k)}$. The specific enthalpy of moist air with temperature $T$ and humidity ratio $W$ is given by \cite{ASHRAE_handbook_fund:17}: $	h(T,W) = C_{pa}T + W(g_{H_20}+C_{pw}T)$,
where $g_{H_20}$ is the heat of evaporation of water at 0~$\degree C$, and $C_{pa},C_{pw}$ are specific heat of air and water at constant pressure.

The reheating coil power consumption in the $i^{th}$ VAV box is modeled to be proportional to the heat it adds to the conditioned air stream:
\begin{align} \label{P_reheat}
	P_{reheat,i}(k) &= \frac{m_{sa,i}(k)C_{pa}\big[T_{sa,i}(k)-T_{ca}(k)\big]}{\eta_{reheat}COP_{h}},
\end{align}
where $\eta_{reheat}$ is the reheating coil efficiency, and $COP_h$ is the boiler coefficient of performance.

\paragraph{Overall Plant}
The overall plant, i.e., virtual building---consisting of the building thermal model, cooling and dehumidifying coil model, and power consumption models---is simulated using SIMULINK and MATLAB\copyright. The building thermal model is constructed in DYMOLA 2021 and it is exported into an FMU (Functional Mockup Unit). It is then imported into SIMULINK using the FMI Kit for SIMULINK. The remaining models are constructed directly in SIMULINK.

\section{Proposed Multi-Zone Hierarchical Control~(\slmpc)}\label{sec:mpc}
\begin{figure}[b]
	\centering
	\includegraphics[width=0.99\linewidth]{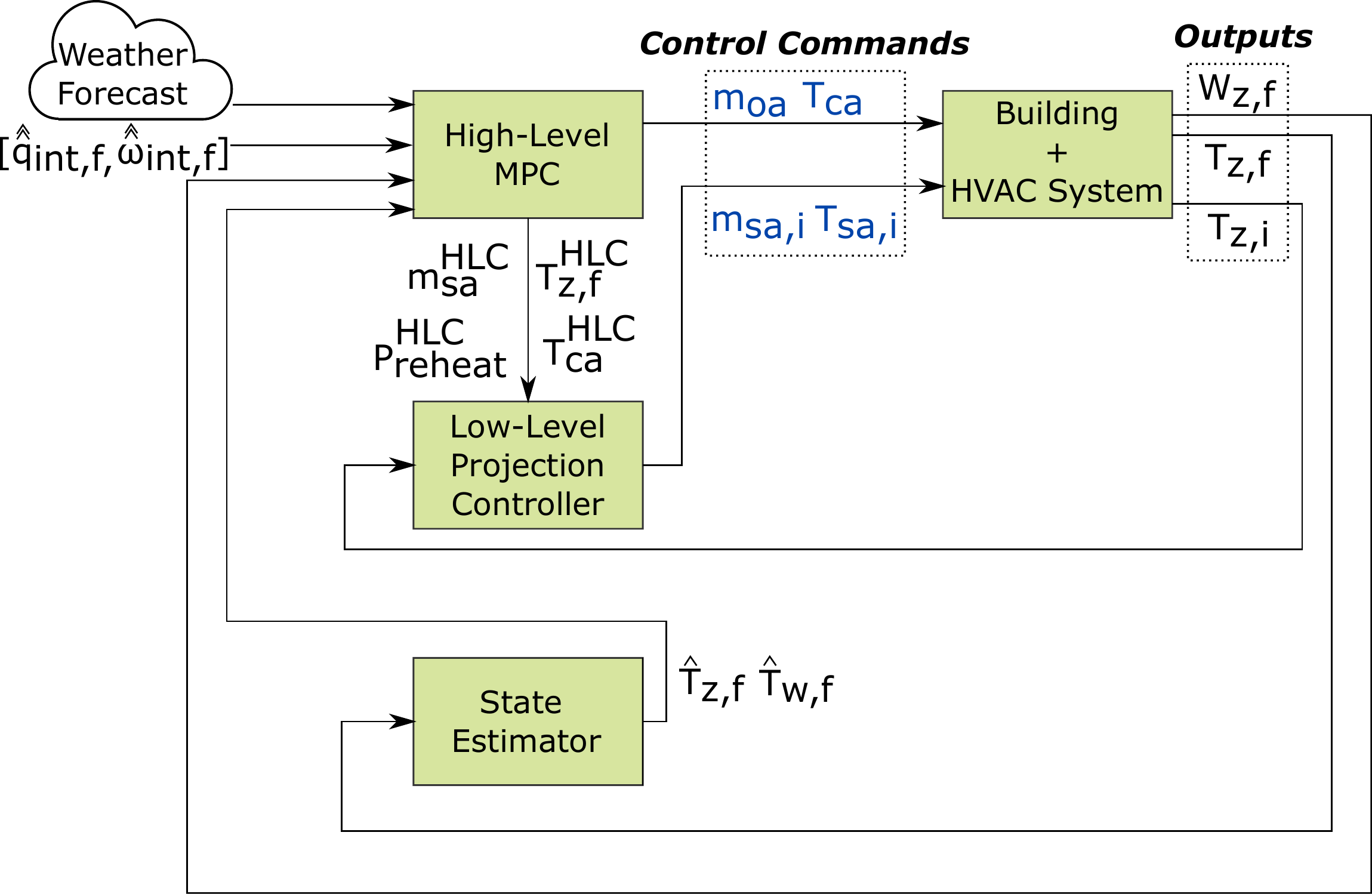}
	\caption{Structure of the proposed multi-zone hierarchical controller (\slmpc{}). We denote estimates as $\hat{\bullet}$ and forecasts as $\hat{\hat{\bullet}}$. Variables with a subscript $i$ are for the individual zones, while the variables with a subscript $f$ represent the aggregate quantities for each floor/meta-zone. In this figure, $T_{z,f}$, $W_{z,f}$, $q_{ac,f}$, $\hat T_{z,f}$, $\hat T_{w,f}$, $\hat{\hat{q}}_{int,f}$, $\hat{\hat{\omega}}_{int,f}$, and $T_{z,f}^{HLC}$ are $\forall f\in \mathbf{F}$; $m_{sa,i}$ and $T_{z,i}$ are for $i\in\mathbf{I_f}, \,\forall \text{f}\in\mathbf{F}$; $T_{sa,i}$ is for $i\in\mathbf{I_{rh,f}}, \,\forall \text{f}\in\mathbf{F}$.}\vspace{-10pt}
	\label{fig:MPC_schematic_multi_zone}
\end{figure}

Recall that both the proposed and the baseline controllers need to decide the following control commands:
\begin{align*}
	u(k) &\eqdef [m_{oa}(k), T_{ca}(k), m_{sa,i}(k), T_{sa,i}(k)]^T\in\Re^{2+n_z+n_z^{rh}}.
\end{align*}
Figure~\ref{fig:MPC_schematic_multi_zone} shows the structure of the proposed \slmpc{}. The high-level controller is based on MPC and decides the control commands for the AHU: outdoor air flow rate~($m_{oa}$) and conditioned air temperature~($T_{ca}$). The low-level controller is a projection-based feedback controller and decides the control commands for each of the VAV boxes/zones: supply air flow rate~($m_{sa,i}$) and supply air temperature~($T_{sa,i}$). These controllers are described in detail next.

\subsection{MPC-Based High-Level Controller (\hlc{})}\label{sec:HLC}
The high-level controller~(\hlc{}) is based on MPC that uses a \loresmodel{} model of the building which is divided into a small number of meta-zones. Each  meta-zone is an aggregation of multiple zones in the real building. This aggregation can be done in any number of ways. In this work we aggregate all the zones in a floor into a meta-zone, which is denoted by $f\in\mathbf{F}\eqdef\{1,\dots,n_f\}$, where $n_f$ is the total number of floors/meta-zones. The Innovation Hub building has three floors, so we aggregate it into three meta-zones. The set of all VAVs/zones in floor $f$ is denoted as $\mathbf{I_f}$ (so $|\cup_{\text{f}\in \mathbf{F}}\mathbf{I_f}|=n_z$), of which those equipped with reheat coils is denoted as $\mathbf{I_{rh,f}}$ (so $|\cup_{\text{f}\in \mathbf{F}}\mathbf{I_{rh,f}}|=n_z^{rh}$). The \hlc{} decides on the following control commands based on the aggregate models:
\begin{align}\label{eq:u_hlc}
	u^{HLC}(k)&\eqdef\big(m_{oa}(k), T_{ca}(k), m_{sa,f}(k), T_{sa,f}(k),\nonumber\\
	&\qquad \qquad \qquad \qquad \forall f\in\mathbf{F}\big)\in \Re^{2+(2\times n_f)},
\end{align}
where $m_{sa,f}(k) \eqdef \sum\limits_{i\in \mathbf{I_f}}m_{sa,i}(k)$ is the aggregate (total) supply airflow rate to all the zones in floor/\metazone{} $f$ and $T_{sa,f}(k)$ is the aggregate supply air temperature. Of the control commands computed in~\eqref{eq:u_hlc}, $m_{oa}(k)$ and $T_{ca}(k)$ can be directly sent to the plant.  The remaining information computed by the \hlc{} including $m_{sa,f}(k)$ and $T_{sa,f}(k)$ are used by the low-level controller (\llc{}), described in Section~\ref{sec:llc}, to decide on the supply airflow rate ($m_{sa,i}(k)$) and supply air temperature ($T_{sa,i}(k)$) for the individual zones/VAV boxes in each floor.

A comment on notation: all variables with a subscript $i$ are for the individual zones, while the variables with a subscript $f$ represent the aggregate quantities for each meta-zone.

For MPC formulation, we use a model interval of $\Delta t=5$~minutes, a control interval of $\Delta T=15$~minutes, and a prediction/planning interval of $T=24$~hours. So we have $T=N\Delta T$ and $\Delta T = M\Delta t$, where $N=96$ (planning horizon) and $M=3$. The control inputs for $N$ time steps are obtained by solving an optimization problem of minimizing the energy consumption subject to thermal comfort, indoor air quality, and actuator constraints. Then the control commands obtained for the first time step are sent to the plant and the \llc{}. The optimization problem is solved again for the next $N$ time steps with the initial states of the model obtained from a state estimator, which uses measurements from the plant. This process is repeated at the next control time step, i.e., after an interval of $\Delta T$. 

To describe the optimization problem, first we define the state vector $x(k)$ and the vector of control commands and internal variables $v(k)$ as:
\begin{align}
	x(k)&\eqdef \big(T_{z,f}(k), T_{w,f}(k), W_{z,f}(k),\,\,\forall f\in \mathbf{F}\big)\in\Re^{3\times n_f}, \label{eq:hlc_states}\\
	v(k)&\eqdef \big(u^{HLC}(k),m_w(k),W_{ca}(k)\big)\in \Re^{2+(2\times n_f)+2}\label{eq:hlc_controls},
\end{align}
where $T_{z,f}(k)$, $T_{w,f}(k)$, and $W_{z,f}(k)$ are the aggregate zone temperature, wall temperature, and humidity ratio of floor/\metazone{} $f$, respectively; $u^{HLC}(k)$ is the control command vector defined in~\eqref{eq:u_hlc} and $m_{w}(k)$ is the chilled water flow rate into the cooling coil. The exogenous input vector is:
\begin{align}\label{eq:hlc_exogenous_inputs}
	w(k)&\eqdef \big(\eta_{sol}(k), T_{oa}(k), W_{oa}(k), q_{int,f}(k), \omega_{int,f}(k),\nonumber \\
	&\qquad \qquad \qquad \qquad \qquad \quad \forall f\in\mathbf{F}\big)\in\Re^{3+(2\times n_f)},
\end{align}
where $\eta_{sol}(k)$ is the solar irradiance, $T_{oa}(k)$ is the outdoor air temperature, $W_{oa}(k)$ is the outdoor air humidity ratio, $q_{int,f}(k)$ is the aggregate internal heat load in floor/\metazone{} $f$ due to occupants, lights, equipment, etc., and $\omega_{int,f}(k)$ is the aggregate rate of water vapor generation in floor/\metazone{} $f$ due to occupants and other sources. We denote the forecast of these exogenous inputs as $\hat{\hat{w}}$; in Section~\ref{sec:simulation_setup}, we discuss how these forecasts are obtained. The vector of nonnegative slack variables $\zeta(k)\eqdef \big(\zeta_{T,f}^{low}(k), \zeta_{T,f}^{high}(k), \zeta_{W,f}^{low}(k), \zeta_{W,f}^{high}(k),\,\,\forall f \in \mathbf{F} \big)\in \Re^{4\times n_f},$ is introduced for feasibility of the optimization problem.

The optimization problem at time index $j$ is:
\begin{subequations}\allowdisplaybreaks
	\begin{align}\label{eq:cost_hlc}
		\min_{V,X,Z} &\sum\limits_{k=j}^{j+NM-1} \bigg[P_{fan}(k) + P_{cc}(k) + \sum\limits_{f\in\mathbf{F}}P_{reheat,f}(k)\bigg] \Delta t + P_{slack}(k),
	\end{align} 
	where $P_{fan}(k)$ is given by \eqref{eq:P_fan}, $P_{cc}(k)$ is given by \eqref{eq:P_cc}, $P_{reheat,f}(k)\eqdef \frac{m_{sa,f}(k)C_{pa}\big[T_{sa,f}(k)-T_{ca}(k)\big]}{\eta_{reheat}COP_{h}}$, $V \eqdef [v^T(j), v^T(j+1), \dots, v^T(j+NM-1)]^T$, $X \eqdef [x^T(j+1), x^T(j+2), \dots, x^T(j+NM)]^T$, and $Z \eqdef [\zeta^T(j+1), \zeta^T(j+2), \dots \zeta^T(j+NM)]^T$. The last term, $P_{slack}$, penalizes the aggregate zone temperature and humidity slack variables:
	\begin{align*}
		P_{slack}(k) \eqdef& \sum\limits_{f\in\mathbf{F}}\bigg[ \lambda_T^{low}\zeta_{T,f}^{low}(k+1) + \lambda_T^{high}\zeta_{T,f}^{high}(k+1) + \\ 
		&\quad\lambda_W^{low}\zeta_{W,f}^{low}(k+1) + \lambda_W^{high}\zeta_{W,f}^{high}(k+1)\bigg],
	\end{align*}
	where the $\lambda$s are penalty parameters. The total supply airflow rate $m_{sa}(k)$ used in $P_{fan}(k)$ and $P_{cc}(k)$, is given by $m_{sa}(k)=\sum\limits_{f\in\mathbf{F}} m_{sa,f}(k)=\sum\limits_{\text{f}\in\mathbf{F}}\sum\limits_{i\in\mathbf{I_f}} m_{sa,i}(k)$. The optimal control commands are obtained by solving the optimization problem~\eqref{eq:cost_hlc} subject to the following constraints:
	\begin{align}\allowdisplaybreaks
		&T_{z,f}(k+1) = T_{z,f}(k) + \Delta t \bigg[ \frac{T_{oa}(k) - T_{z,f}(k)}{\tau_{za,f}} + \nonumber \\
		&\,\,\,\frac{T_{w,f}(k) - T_{z,f}(k)}{\tau_{zw,f}} + A_{z,f} \eta_{sol}(k) + \frac{q_{int,f}(k) + q_{ac,f}(k)}{C_{z,f}} \bigg]\label{eq:hlc_T_z}\\[0.25em]
		&T_{w,f}(k+1) = T_{w,f}(k) + \Delta t \bigg[ \frac{T_{oa}(k) - T_{w,f}(k)}{\tau_{wa,f}} 
		+ \nonumber \\
		&\qquad \qquad \qquad \qquad  \frac{T_{z,f}(k) - T_{w,f}(k)}{\tau_{wz,f}} + A_{w,f} \eta_{sol}(k) \bigg]\label{eq:hlc_T_w}\\[0.25em]
		&W_{z,f}(k+1) = W_{z,f}(k)+\frac{\Delta t R_gT_{z,f}(k)}{V_fP^{da}}\bigg[\omega_{int,f}(k) + \nonumber \\
		&\qquad \qquad \qquad \qquad \qquad m_{sa,f}(k)\frac{W_{ca}(k)-W_{z,f}(k)}{1+W_{ca}(k)}\bigg]\label{eq:hlc_W_z}\\[0.25em]
		&T_{ca}(k) = T_{ma}(k) + m_w(k)f\big(T_{ma}(k),W_{ma}(k),m_{sa}(k),m_{w}(k)\big)\label{eq:hlc_T_cc}\\[0.25em]
		&W_{ca}(k) = W_{ma}(k) + m_w(k)g\big(T_{ma}(k),W_{ma}(k),m_{sa}(k),m_{w}(k)\big)\label{eq:hlc_W_cc}\\[0.25em]
		&T_{z,f}^{low}(k)-\zeta_{T,f}^{low}(k) \leq T_{z,f}(k) \leq T_{z,f}^{high}(k) + \zeta_{T,f}^{high}(k) \label{eq:hlc_T_z_box}\\[0.25em]
		&a^{low}T_{z,f}(k) + b^{low} -\zeta_{W,f}^{low}(k) \leq W_{z,f}(k) \nonumber \\
		&\qquad \qquad \qquad \qquad \leq a^{high}T_{z,f}(k) + b^{high} +\zeta_{W,f}^{high}(k)\label{eq:hlc_W_z_box}\\[0.25em]
		&m_{oa}^{min} \leq m_{oa}(k)\leq m_{oa}^{max}\label{eq:hlc_m_oa}\\[0.25em]
		&T_{ca}(k+1) \leq min\big(T_{ca}(k)+T_{ca}^{rate}\Delta t,T_{ma}(k+1),T_{ca}^{high}\big)\label{eq:hlc_T_ca_high}\\[0.25em]
		&T_{ca}(k+1) \geq max\big(T_{ca}(k)-T_{ca}^{rate}\Delta t,T_{ca}^{low}\big)\label{eq:hlc_T_ca_low}\\[0.25em]
		&W_{ca}(k) \leq W_{ma}(k)\label{eq:hlc_W_ca}\\[0.25em]
		&r_{oa}(k) = m_{oa}(k)/m_{sa}(k)\label{eq:hlc_r_oa_def}\\[0.25em]
		&r_{oa}(k+1) \leq min\big(r_{oa}(k)+r_{oa}^{rate}\Delta t,r_{oa}^{high}\big)\label{eq:hlc_r_oa_high}\\[0.25em]
		&r_{oa}(k+1) \geq max\big(r_{oa}(k)-r_{oa}^{rate}\Delta t,r_{oa}^{low}\big)\label{eq:hlc_r_oa_low}\\[0.25em]
		&m_{sa,f}^{low} \leq m_{sa,f}(k) \leq m_{sa,f}^{high}\label{eq:hlc_m_sa}\\[0.25em]
		&T_{ca}(k) \leq T_{sa,f}(k) \leq T_{sa}^{high}\label{eq:hlc_T_sa}\\[0.25em]
		&\zeta_{T,f}^{low}(k+1), \zeta_{T,f}^{high}(k+1), \zeta_{W,f}^{low}(k+1), \zeta_{W,f}^{high}(k+1)\geq 0\label{eq:hlc_slack}
	\end{align}
\end{subequations}
where constraints~\eqref{eq:hlc_T_z}-\eqref{eq:hlc_W_z}, \eqref{eq:hlc_T_z_box}-\eqref{eq:hlc_W_z_box}, and \eqref{eq:hlc_m_sa}-\eqref{eq:hlc_slack} are $\forall f\in \mathbf{F}$. Constraints~\eqref{eq:hlc_T_z}-\eqref{eq:hlc_W_cc}, \eqref{eq:hlc_m_oa}, \eqref{eq:hlc_W_ca}-\eqref{eq:hlc_r_oa_def}, and \eqref{eq:hlc_m_sa}-\eqref{eq:hlc_slack} are for $k\in\{j,\dots,j+NM-1\}$, constraints \eqref{eq:hlc_T_z_box} and \eqref{eq:hlc_W_z_box} are for $k\in\{j+1,\dots,j+NM\}$, and constraints \eqref{eq:hlc_T_ca_high}-\eqref{eq:hlc_T_ca_low} and \eqref{eq:hlc_r_oa_high}-\eqref{eq:hlc_r_oa_low} are for $k\in\{j-1,\dots,j+NM-2\}$. The control commands remain the same for $M$ time steps, as the control interval $\Delta T= M\Delta t$, i.e., $u^{HLC}(k)=u^{HLC}(k+1)=\dots=u^{HLC}(k+M-1),\,\, \forall k\in \{j,j+M,\dots,j+NM-1\}$.

Constraints~\eqref{eq:hlc_T_z} and \eqref{eq:hlc_T_w} are due to the aggregate thermal dynamics of floor/\metazone{} $f$, which is a discretized form of an RC (resistor-capacitor) network model, specifically a 2R2C model. The two states of the model are aggregate zone temperature ($T_{z,f}$) and wall temperature ($T_{w,f}$, a fictitious state). In constraint~\eqref{eq:hlc_T_z}, $q_{ac,f}(k)$ is the heat influx due to the HVAC system and is given by  $q_{ac,f}(k) \eqdef m_{sa,f}(k)C_{pa}(T_{sa,f}(k) - T_{z,f}(k))$. The model has seven parameters $\{C_{z,f}, \tau_{zw,f}, \tau_{za,f}, A_{z,f}, \tau_{wz,f}, \tau_{wa,f}, A_{w,f}\}$. In the evaluation study, they are estimated using the algorithm presented in~\cite{GuoAggregationEnB:2020} and will be discussed later in Section~\ref{sec:simulation_setup}.

The constraint~\eqref{eq:hlc_W_z} is for the aggregate humidity dynamics of floor/\metazone{} $f$, where $W_{z,f}$ is the aggregate zone humidity ratio, $V_{f}$ is the volume of \metazone{} $f$, $R_g$ is the specific gas constant of dry air, $P^{da}$ is the partial pressure of dry air, and $W_{ca}$ is the conditioned air humidity ratio~\cite{SG_PB_Energy:11}. 

Constraints~\eqref{eq:hlc_T_cc} and \eqref{eq:hlc_W_cc} are for the control-oriented cooling and dehumidifying coil model, which was developed in our prior work~\cite{RamanMPC_AE:2020}. The specific functional form in~\eqref{eq:hlc_T_cc} and \eqref{eq:hlc_W_cc} is chosen so that when the chilled water flow rate is zero, no cooling or dehumidification of the air occurs, so the conditioned air temperature and humidity ratio are equal to the mixed air temperature and humidity ratio: $T_{ca}=T_{ma}$ and $W_{ca}=W_{ma}$, when $m_w=0$. The interested readers are referred to~\cite[Section 3.1.1]{RamanMPC_AE:2020} for details regarding the model. 

Constraints~\eqref{eq:hlc_T_z_box} and \eqref{eq:hlc_W_z_box} are box constraints to maintain the temperature and humidity of the \metazones{} within the allowed comfort limits. The constraints are softened using slack variables $\zeta_{T,f}^{low}(k)$, $\zeta_{T,f}^{high}(k)$, $\zeta_{W,f}^{low}(k)$, and $\zeta_{W,f}^{high}(k)$; constraint~\eqref{eq:hlc_slack} ensures that these slack variables are nonnegative. Imposing constraints directly on the relative humidity of zones ($RH_z$) is difficult, as relative humidity is a highly nonlinear function of dry bulb temperature and humidity ratio~\cite[Chapter~1]{ASHRAE_handbook_fund:17}. So we linearize this function which gives us $a^{low}$, $b^{low}$, $a^{high}$, and $b^{high}$ in \eqref{eq:hlc_W_z_box}, and helps in converting the constraints on relative humidity to humidity ratio ($W_z$).

Constraint~\eqref{eq:hlc_m_oa} is for the outdoor airflow rate, where the minimum allowed value ($m_{oa}^{min}$) is computed based on the ventilation requirements specified in ASHRAE 62.1~\cite{ASHRAE62-2016} and to maintain positive building pressurization.

Constraints~\eqref{eq:hlc_T_ca_high}-\eqref{eq:hlc_T_ca_low} and \eqref{eq:hlc_r_oa_high}-\eqref{eq:hlc_r_oa_low} are to take into account the capabilities of the cooling coil and damper actuators. In constraints~\eqref{eq:hlc_T_ca_high} and \eqref{eq:hlc_W_ca}, the inequalities $T_{ca}(k+1)\leq T_{ma}(k+1)$ and $W_{ca}(k)\leq W_{ma}(k)$ ensure that the cooling coil can only cool and dehumidify the mixed air stream; it cannot add heat or moisture. Similarly, in constraint~\eqref{eq:hlc_T_sa} the inequality $T_{sa,f}(k)\geq T_{ca}(k)$ ensures that the reheat coils can only add heat; it cannot cool.

Constraint~\eqref{eq:hlc_m_sa} is to take into account the capabilities of the fan and aggregate capabilities of the VAV boxes. The limits $m_{sa,f}^{low}$ and $m_{sa,f}^{high}$ are computed using the VAV schedule from the mechanical drawings of a building as follows: $m_{sa,f}^{low} \eqdef \sum\limits_{i\in \mathbf{I_{f}}}m_{sa,i}^{low}$ and $m_{sa,f}^{high} \eqdef \sum\limits_{i\in \mathbf{I_{f}}}m_{sa,i}^{high}$.

Note that of the states $x(k)$ defined in~\eqref{eq:hlc_states}, $T_{w,f}$ is a fictitious state that cannot be measured, while the other two states aggregate zone temperature ($T_{z,f}$) and aggregate zone humidity ratio ($W_{z,f}$) are measured. So we estimate the current state $\hat x(k)$ using a Kalman filter.

\subsection{Projection-Based Low-Level Controller (\llc{})}\label{sec:llc}
The role of the low-level controller (\llc{}) is to appropriately distribute the aggregate quantities---such as the total supply airflow rate and reheat power consumption---computed by the \hlc{} to individual zones/VAVs. The \llc{} needs to do so by capturing two important properties: (i)~it should consider the needs of individual zones and distribute accordingly, and (ii)~it should act in coherence with the \hlc{}, so that there is minimal mismatch for the MPC optimization in the next round.

The \llc{} is a projection-based feedback controller that decides on the supply airflow rate and supply air temperature for each VAV box/zone. That is, the control command vector that the \llc{} needs to decide is:
\begin{align*}
	u^{LLC}(k) &\eqdef [m_{sa,i}(k), T_{sa,i}(k)]^T\in\Re^{n_z+n_z^{rh}}, 
\end{align*}
where for $m_{sa,i}$, $i \in \mathbf{I_f}, \,\,\forall \text{f} \in \mathbf{F}$, and for $T_{sa,i}$, $i\in \mathbf{I_{rh,f}},\,\, \forall \text{f} \in \mathbf{F}$. It decides these control commands by using the following information from the \hlc{}: (i)~total allowed supply airflow rate to all the zones $m_{sa}(k)=\sum\limits_{f\in\mathbf{F}}m_{sa,f}(k)$, (ii)~total allowed reheat power consumption $P_{reheat}(k) = \sum\limits_{f\in\mathbf{F}}P_{reheat,f}(k)$, (iii)~the temperature at which the zones in each \metazone{} should be maintained at~$T_{z,f}(k+1)$, and (iv) the conditioned air temperature $T_{ca}(k)$. Here on in this section, we will be using the superscript \hlc{} ($\bullet^{HLC}$) for these variables to make it clear that these are obtained from the high-level controller.

First the needs of each zone are assessed based on the current measured temperature $T_{z,i}(k)$ and the range it should be in~$[T_{z,i}^{htg}(k), T_{z,i}^{clg}(k)]$ and are translated into the desired supply airflow rate $m_{sa,i}^d(k)$ and supply air temperature $T_{sa,i}^d(k)$. 
Then these desired values along with the information obtained from the \hlc{} are used to solve a projection problem to compute the control commands for all the zones, $u^{LLC}(k)$. 

\begin{figure}[b]
	\centering
	\includegraphics[width=0.99\linewidth]{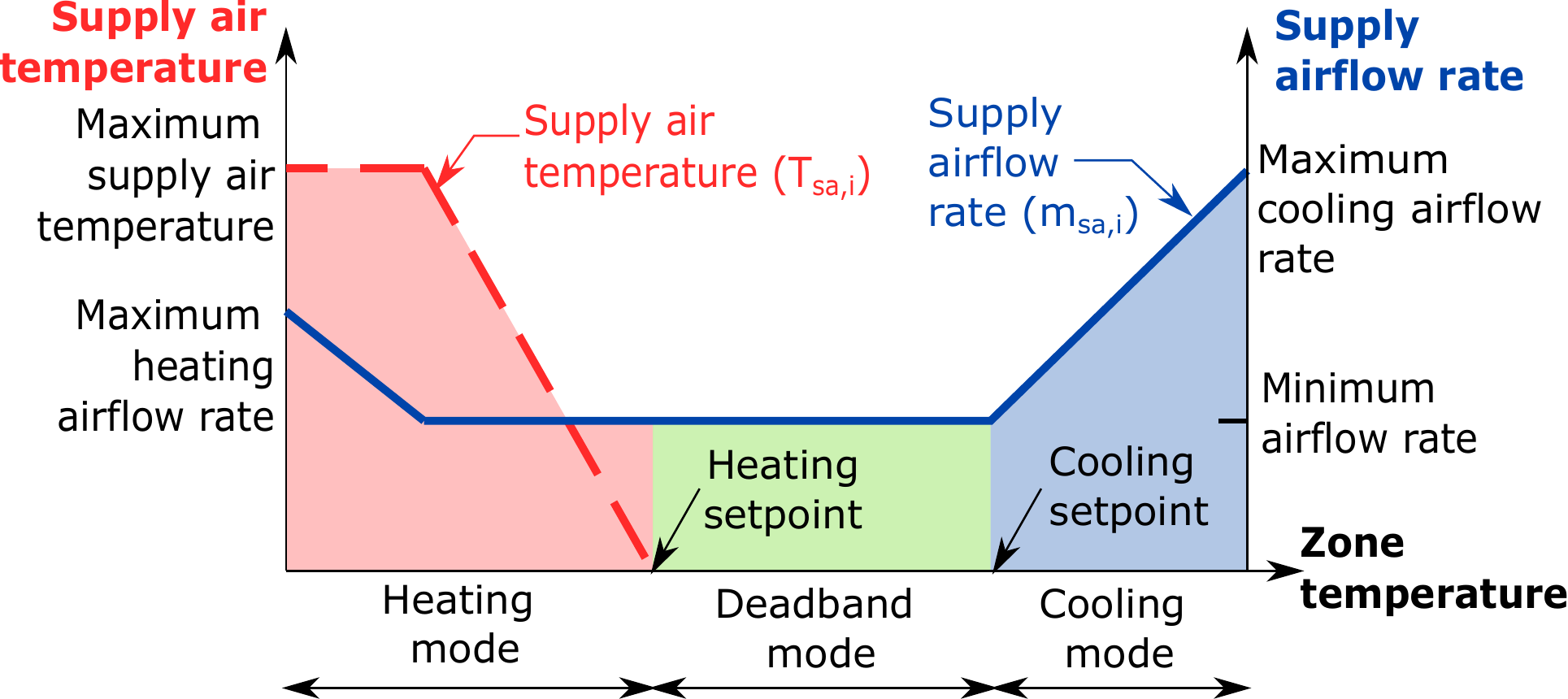}
	\caption{Schematic of the \emph{Dual Maximum} control algorithm.}
	\label{fig:dual_maximum_controller}
\end{figure}

The procedure used to compute the desired values $m_{sa,i}^d(k)$ and $T_{sa,i}^d(k)$ is explained below. This is similar to the \emph{Dual Maximum} control logic presented in Section~\ref{sec:baseline}; a schematic representation of it is shown in Figure~\ref{fig:dual_maximum_controller}.
\begin{enumerate}
	\item First the temperature range $[T_{z,i}^{htg}(k),\, T_{z,i}^{clg}(k)]$ in which each zone should be is computed as follows: $T_{z,i}^{htg}(k) = max\big(T_{z,f}^{HLC}(k+1)-\tilde T_z^{db}/2,\,T_{z,f}^{low}\big)\,\, \forall i\in \mathbf{I_f}$ and $T_{z,i}^{clg}(k) = min\big(T_{z,f}^{HLC}(k+1)+\tilde T_z^{db}/2,\,T_{z,f}^{high}\big)\,\, \forall i\in \mathbf{I_f}$, where $T_{z,f}^{HLC}(k+1)$ is obtained from the \hlc{}, $\tilde T_z^{db}$ is a deadband, and $T_{z,f}^{low}$ and $T_{z,f}^{high}$ are the limits used in constraint~\eqref{eq:hlc_T_z_box}.
	\item If the zone temperature is between the cooling and heating setpoints ($T_{z,i}(k)\in[T_{z,i}^{htg}(k),\, T_{z,i}^{clg}(k)]$), then the controller is in deadband mode. The supply airflow rate is desired to be at its minimum and no heating is required, i.e., $m_{sa,i}^d(k)=m_{sa,i}^{low}$ and $T_{sa,i}^d(k)=T_{ca}^{HLC}(k)$. 
	\item If the zone temperature is warmer than the cooling setpoint ($T_{z,i}(k)>T_{z,i}^{clg}(k)$), then the controller is in cooling mode. The supply airflow rate is desired to be increased as needed and no heating is required, i.e., $m_{sa,i}^d(k)=min\big(m_{sa,i}^{low} + K_{m,i}^{clg}(T_{z,i}(k)-T_{z,i}^{clg}(k)),\, m_{sa,i}^{high}\big)$ and $T_{sa,i}^d(k)=T_{ca}^{HLC}(k)$.
	\item If the zone temperature is cooler than the heating setpoint ($T_{z,i}(k)<T_{z,i}^{htg}(k)$), then the controller is in heating mode. Heating is required, and the supply airflow rate is desired to be increased only if additional heating is needed, i.e., $T_{sa,i}^d(k)=min\big(T_{ca}^{HLC}(k)+K_{T,i}^{htg}(T_{z,i}^{htg}(k)-T_{z,i}(k)),\,T_{sa}^{high}\big)$;  if $T_{sa,i}^d(k)=T_{sa}^{high}$, then $m_{sa,i}^d(k)=min\big(m_{sa,i}^{low} + K_{m,i}^{htg}(T_{z,i}^{htg}(k)-T_{z,i}(k)),\, m_{sa,i}^{high,reheat}\big)$, otherwise $m_{sa,i}^d(k)=m_{sa,i}^{low}$.
	\item Finally, we impose the following rate constraints: 
	\begingroup
	\begin{align*}
		&m_{sa,i}(k-M)-m_{sa,i}^{rate}\Delta T\leq m_{sa,i}^d(k)\\
		&\qquad \qquad \qquad \qquad \qquad \leq m_{sa,i}(k-M)+m_{sa,i}^{rate}\Delta T, \nonumber\\
		&T_{sa,i}(k-M)-T_{sa,i}^{rate}\Delta T\leq T_{sa,i}^d(k)\\
		&\qquad \qquad \qquad \qquad \qquad \leq T_{sa,i}(k-M)+T_{sa,i}^{rate}\Delta T, 
	\end{align*}
	\endgroup
	where $m_{sa,i}(k-M)$ and $T_{sa,i}(k-M)$ are the supply airflow rate and supply air temperature from the previous control time step.
\end{enumerate}

These desired values---$m_{sa,i}^d(k)$ and $T_{sa,i}^d(k)$---along with information from the \hlc{} are used to solve the following projection problem to obtain the control commands for all the zones,~$u^{LLC}(k)$:
\begin{subequations}
	\begin{align}
		\min_{u^{LLC}(k)} &\sum\limits_{\text{f}\in\mathbf{F}}\sum_{i\in\mathbf{I_f}}\lambda_{m,i}(m_{sa,i}(k)-m_{sa,i}^d(k))^2 + \nonumber \\ 
		&\qquad \qquad  \sum\limits_{\text{f}\in\mathbf{F}}\sum_{i\in\mathbf{I_{rh,f}}}\lambda_{T,i}(T_{sa,i}(k)-T_{sa,i}^d(k))^2
	\end{align}
	subject to the following constraints:
	\begin{align}
		&\sum\limits_{\text{f}\in\mathbf{F}}\sum_{i\in\mathbf{I_f}}m_{sa,i}(k)\leq m_{sa}^{HLC}(k) \label{eq:llc_m_sa_hlc}\\
		&\sum\limits_{\text{f}\in\mathbf{F}}\sum_{i\in\mathbf{I_{rh,f}}} \frac{m_{sa,i}(k)C_{pa}\Big(T_{sa,i}(k)-T_{ca}^{HLC}(k)\Big)}{\eta_{reheat}COP_h}\leq P_{reheat}^{HLC}(k) \label{eq:llc_P_reheat_hlc}\\
		&m_{sa,i}^{low}\leq m_{sa,i}(k) \leq m_{sa,i}^{high}, \quad\forall i\in\mathbf{I_f},\,\, \forall \text{f}\in\mathbf{F} \label{eq:llc_m_sa}\\
		&T_{ca}(k)\leq T_{sa,i}(k) \leq T_{sa}^{high}, \quad\forall i\in\mathbf{I_{rh,f}},\,\, \forall \text{f}\in\mathbf{F} \label{eq:llc_T_sa}
	\end{align}
\end{subequations}
where the sets $\mathbf{I_f}$ and $\mathbf{I_{rh,f}}$ are defined at the beginning of this section, $\lambda$s are weights, $m_{sa}^{HLC}(k)=\sum\limits_{f\in\mathbf{F}}m_{sa,f}^{HLC}(k)$, and $P_{reheat}^{HLC}(k) = \sum\limits_{f\in\mathbf{F}}P_{reheat,f}^{HLC}(k)$.

Constraints~\eqref{eq:llc_m_sa_hlc} and \eqref{eq:llc_P_reheat_hlc} are to ensure that the total supply airflow rate and reheat power consumption do not exceed the limits computed by the \hlc{}. Constraints~\eqref{eq:llc_m_sa} and \eqref{eq:llc_T_sa} are to take in to account the capabilities of the VAV boxes and reheat coils. In constraint~\eqref{eq:llc_T_sa}, the inequality $T_{sa,i}(k)\geq T_{ca}(k)$ ensures that the reheat coils can only add heat to the conditioned air and cannot cool. The upper limit on supply air temperature ($T_{sa}^{high}$) in constraint~\eqref{eq:llc_T_sa} is to prevent thermal stratification~\cite{ASHRAE_handbook_applications:11}.

\section{Baseline Control (\bl)}\label{sec:baseline}
For zone climate control, we consider the \emph{Dual Maximum} algorithm~\cite{ASHRAE_handbook_applications:11} as the baseline; a schematic representation of this algorithm is shown in Figure~\ref{fig:dual_maximum_controller}. Even though \emph{Single Maximum} is more commonly used, including in the Innovation Hub building, we choose \emph{Dual Maximum} for the baseline, as it is more energy-efficient among the two~\cite{ASHRAE_handbook_applications:11,SG_HI_PB_AE:2013}. The \emph{Dual Maximum} controller operates in three modes based on the zone temperature ($T_{z,i}$): (i)~cooling, (ii)~heating, and (iii)~deadband.
\ifArxivVersion
The zone's supply airflow rate ($m_{sa,i}$) and supply air temperature ($T_{sa,i}$) are varied based on the mode, as explained below.
\begin{enumerate}
	\item \textbf{Cooling mode:} If the zone temperature is warmer than the cooling setpoint, then the controller is in cooling mode. The supply airflow rate ($m_{sa,i}$) is varied between the minimum~($m_{sa,i}^{low}$) and maximum~($m_{sa,i}^{high}$) as needed, and the supply air temperature ($T_{sa,i}$) is equal to the conditioned air temperature ($T_{ca}$), i.e., no reheat.
	\item \textbf{Heating mode:} If the zone temperature is below the heating setpoint, then the controller is in heating mode. First, the supply air temperature ($T_{sa,i}$) is increased up to the maximum allowed value~($T_{sa}^{high}$) as needed to maintain the zone temperature at the heating setpoint. If the zone temperature still cannot be maintained at the heating setpoint, then the supply airflow rate is increased between the minimum~($m_{sa,i}^{low}$) and the heating maximum~($m_{sa,i}^{high,reheat}$) values.
	\item \textbf{Deadband mode:} If the zone temperature is between the heating and cooling setpoints, then the controller is in deadband mode. The supply airflow rate is kept at the minimum, and the supply air temperature is equal to the conditioned air temperature, i.e., no reheat.
\end{enumerate}
In the case of VAV boxes that do not have reheat coils, the logic during cooling and deadband modes are the same. In heating mode, the supply airflow rate is at the minimum and the supply air temperature is equal to the conditioned air temperature, as the VAV box cannot heat.
\else
The zone's supply airflow rate ($m_{sa,i}$) and supply air temperature ($T_{sa,i}$) are varied based on the mode the controller is in; see Figure~\ref{fig:dual_maximum_controller}.
\fi

For the AHU-level commands, the \bl{} controller uses fixed conditioned air temperature that is determined based on expected thermal (sensible and latent) load, and fixed outdoor airflow rates based on ventilation requirements, e.g., ASHRAE~62.1~\cite{ASHRAE62-2016}. Another consideration in choosing outdoor airflow rate is building positive pressurization requirements~\cite{ASHRAE_handbook_fund:17}.

\section{Simulation Setup}\label{sec:simulation_setup}
\ifArxivVersion
\bgroup
\def\arraystretch{1.1}%
\begin{table}[tb]
	\caption{VAV schedule.}
	\label{table:vav_schedule}
	\footnotesize
	\centering
	\begin{tabular}{|c|c|c|c|c|}
		\hline
		\rule{0pt}{4ex} \textbf{VAV} & \textbf{Reheat} & \begin{tabular}[c]{@{}c@{}}$\mathbf{m_{sa,i}^{low}}$\\ $(kg/s)$\end{tabular} & \begin{tabular}[c]{@{}c@{}}$\mathbf{m_{sa,i}^{high,reheat}}$\\ $(kg/s)$\end{tabular} & \begin{tabular}[c]{@{}c@{}}$\mathbf{m_{sa,i}^{high}}$\\ $(kg/s)$\end{tabular} \\ \hline
		101          & Yes             & 0.27                                                                       & 0.68                                                                               & 1.36                                                                        \\ \hline
		102          & Yes             & 0.05                                                                       & 0.10                                                                               & 0.20                                                                        \\ \hline
		103          & Yes             & 0.07                                                                       & 0.17                                                                               & 0.34                                                                        \\ \hline
		104          & Yes             & 0.57                                                                       & 0.57                                                                               & 1.14                                                                        \\ \hline
		105          & Yes             & 0.30                                                                       & 0.30                                                                               & 0.60                                                                        \\ \hline
		106          & No              & 0.04                                                                       & -                                                                                  & 0.23                                                                        \\ \hline
		107          & No              & 0.13                                                                       & -                                                                                  & 0.45                                                                        \\ \hline
		108          & Yes             & 0.14                                                                       & 0.33                                                                               & 0.66                                                                        \\ \hline
		109          & Yes             & 0.08                                                                       & 0.21                                                                               & 0.39                                                                        \\ \hline
		201          & Yes             & 0.21                                                                       & 0.52                                                                               & 1.03                                                                        \\ \hline
		202          & Yes             & 0.11                                                                       & 0.28                                                                               & 0.57                                                                        \\ \hline
		203          & Yes             & 0.11                                                                       & 0.28                                                                               & 0.57                                                                        \\ \hline
		204          & Yes             & 0.06                                                                       & 0.14                                                                               & 0.28                                                                        \\ \hline
		205          & No              & 0.13                                                                       & -                                                                                  & 0.57                                                                        \\ \hline
		206          & Yes             & 0.07                                                                       & 0.16                                                                               & 0.33                                                                        \\ \hline
		207          & Yes             & 0.13                                                                       & 0.32                                                                               & 0.64                                                                        \\ \hline
		208          & Yes             & 0.09                                                                       & 0.20                                                                               & 0.41                                                                        \\ \hline
		209          & Yes             & 0.18                                                                       & 0.18                                                                               & 0.57                                                                        \\ \hline
		210          & Yes             & 0.11                                                                       & 0.28                                                                               & 0.57                                                                        \\ \hline
		211          & Yes             & 0.10                                                                       & 0.26                                                                               & 0.51                                                                        \\ \hline
		212          & Yes             & 0.14                                                                       & 0.34                                                                               & 0.68                                                                        \\ \hline
		301          & Yes             & 0.16                                                                       & 0.41                                                                               & 0.82                                                                        \\ \hline
		302          & Yes             & 0.11                                                                       & 0.28                                                                               & 0.57                                                                        \\ \hline
		303          & Yes             & 0.28                                                                       & 0.28                                                                               & 0.57                                                                        \\ \hline
		304          & Yes             & 0.13                                                                       & 0.32                                                                               & 0.64                                                                        \\ \hline
		305          & Yes             & 0.10                                                                       & 0.23                                                                               & 0.47                                                                        \\ \hline
		306          & Yes             & 0.11                                                                       & 0.28                                                                               & 0.57                                                                        \\ \hline
		307          & No              & 0.13                                                                       & -                                                                                  & 0.57                                                                        \\ \hline
		308          & Yes             & 0.11                                                                       & 0.28                                                                               & 0.57                                                                        \\ \hline
		309          & Yes             & 0.11                                                                       & 0.28                                                                               & 0.57                                                                        \\ \hline
		310          & Yes             & 0.28                                                                       & 0.28                                                                               & 0.57                                                                        \\ \hline
		311          & Yes             & 0.16                                                                       & 0.40                                                                               & 0.80                                                                        \\ \hline
		312          & Yes             & 0.10                                                                       & 0.26                                                                               & 0.51                                                                        \\ \hline
	\end{tabular}
\end{table}
\egroup
\fi

Recall that the plant is based on an air handling unit serving 33 zones, of which 29 are equipped with reheat coils, and the remaining 4 do not have reheat coils (cooling only).
\ifArxivVersion See Table~\ref{table:vav_schedule} and Figure~\ref{fig:floor_plans} for the entire list of VAV boxes/zones. \fi Of the 29 VAV boxes with reheat, three of them serve laboratories which are equipped with fume hoods (209, 303, and 310), and one of them serve restrooms (103). The VAV boxes serving these labs need to be controlled to satisfy the negative pressurization requirements with respect to corridor, so we assume that they operate according to the existing rule based feedback control strategy. Therefore, $n_z=29$ and $n_z^{rh}=25$; for $m_{sa,i}$, $i\in\{ \mathbf{I_1} \eqdef \{101\shyphen102, 104\shyphen109$\}, $\mathbf{I_2}\eqdef\{201\shyphen208, 210\shyphen212\}$, $\mathbf{I_3} \eqdef \{301\shyphen302, 304\shyphen309$, $311\shyphen312\}$ and for $T_{sa,i}$, $i\in\{\mathbf{I_{rh,1}}\eqdef\{\mathbf{I_1}\backslash \{106,107\}\}, \mathbf{I_{rh,2}}\eqdef\{\mathbf{I_2}\backslash \{205\}\}, \mathbf{I_{rh,3}}\eqdef\{\mathbf{I_3}\backslash \{307\}\}\}$. The sets $\mathbf{I_1}$, $\mathbf{I_2}$, and $\mathbf{I_3}$ defined above are the VAVs/zones in floors 1, 2, and 3, respectively. The sets $\mathbf{I_{rh,1}}$, $\mathbf{I_{rh,2}}$, and $\mathbf{I_{rh,3}}$ exclude the VAV boxes which do not have a reheat coil.

The outdoor weather data used in simulations is obtained from the National Solar Radiation Database~\cite{NSRDB} for Gainesville, Florida. As mentioned in Section~\ref{sec:building_thermal_model}, the internal heat load due to occupants are computed based on the number of occupants provided to the zone block. We assume that the zones are occupied from Monday to Friday between 8:00~a.m. to noon and 1:00~p.m. to 5:00~p.m., with the total number of occupants ($n_{p,f}$) in floor~1 as 24, in floor~2 as 26, and in floor~3 as 22. We assume a power density of 12.92~W/m$^2$ (1.2 W/ft$^2$) for internal heat load due to lighting and equipment. For special purpose rooms like electrical and telecommunication, we use a higher power density of 53.82~W/m$^2$ (5 W/ft$^2$). These heat loads from lighting and equipment are assumed to be halved during weekends.
\begin{figure}[tb]
	\centering
	\includegraphics[width=0.99\linewidth]{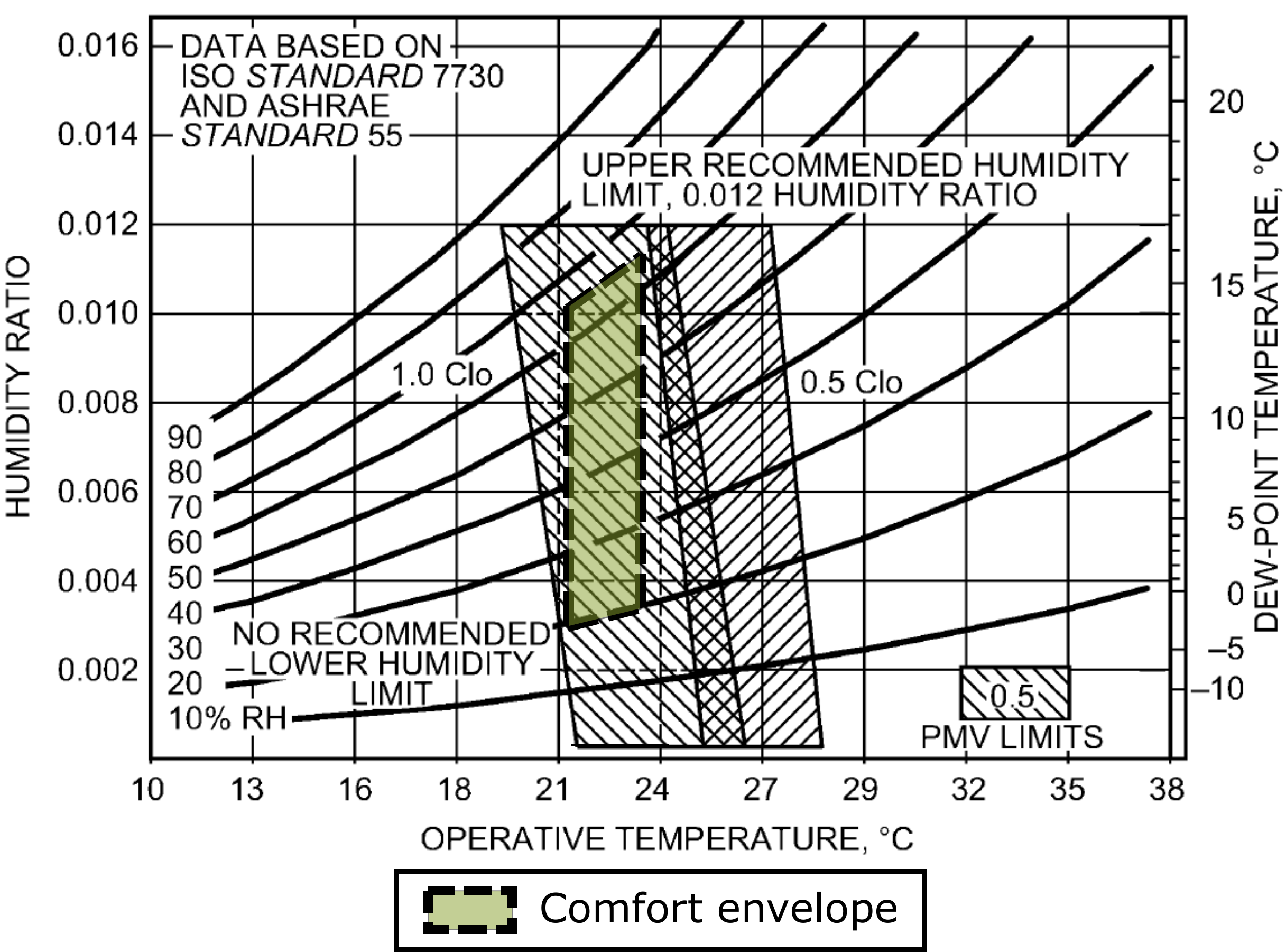}
	\caption{Thermal comfort envelope from \cite{ASHRAE_handbook_fund:17} shown as the hatched areas. Comfort envelope chosen in this paper shown as the green shaded area.}
	\label{fig:thermal_comfort_envelope}\vspace{-10pt}
\end{figure}

The following zone temperature and humidity limits are used in the simulations: $T_{z}^{low}$ = $21.1\degree$C (70$\degree$F), $T_{z}^{high}$ = $23.3\degree$C (74$\degree$F), $RH_{z}^{low}$ = 20\%, and $RH_{z}^{high}$ = 65\%. The chosen thermal comfort envelope is shown in Figure~\ref{fig:thermal_comfort_envelope}. Typically the zone temperature limits during unoccupied mode (unocc) are relaxed when compared to the occupied mode (occ), i.e., $[T_{z}^{low,occ},T_{z}^{high,occ}]\subseteq[T_{z}^{low,unocc},T_{z}^{high,unocc}]$. Due to its usage, the Innovation Hub building is always operated in occupied mode, so we assume the same in simulations. For the simulation results reported later, the zone temperature violation is computed as $max\big(T_z(k)-T_z^{high},\,\, T_z^{low}-T_z(k),\,\,0\big)$ and the zone relative humidity violation is computed as $max\big(RH_z(k)-RH_z^{high},\,\, RH_z^{low}-RH_z(k),\,\,0\big)$, with the upper and lower limits mentioned above.

The fan power coefficient $\alpha_{fan}$ in~\eqref{eq:P_fan} is 14.2005~$W/(kg/s)^3$, which is obtained using a least squares fit to data collected from the building.
The parameters of the cooling and dehumidifying coil model used in the plant are fit using the procedure explained in Section~2.1.2 of \cite{RamanMPC_AE:2020}. The root mean square errors for the validation data set are 0.25$\degree$C (0.46$\degree$F, 2\%) for $T_{ca}$ and 0.22$\times10^{-4}kg_w/kg_{da}$ (2.6\%) for $W_{ca}$. 

\ifArxivVersion The AHU in the building is equipped with a draw-through supply fan and therefore the fan is located after the cooling coil. The fan emits heat, which leads to a slight increase in the conditioned air temperature before it is supplied to the VAV boxes. For the simulations, we assume this increase in temperature to be 1.11$\degree$C (2$\degree$F). \fi

\ifArxivVersion
\bgroup
\def\arraystretch{1.2}%
\begin{table}[tb]
	\caption{Parameters used for the aggregate thermal dynamic model in the \hlc{}.}
	\label{table:thermal_model_mpc}
	\footnotesize
	\centering
	\begin{tabular}{|c|c|c|c|c|}
		\hline
		\textbf{Parameter} & \textbf{Units} & \textbf{Floor 1}     & \textbf{Floor 2}     & \textbf{Floor 3}     \\ \hline
		$C_{z,f}$          & $kWh/^oC$      & 2.9282               & 8.0837               & 8.4974               \\ \hline
		$\tau_{zw,f}$      & hours          & 0.5108               & 2.2161               & 2.5622               \\ \hline
		$\tau_{za,f}$      & hours          & 200                  & 150                  & 150                  \\ \hline
		$A_{z,f}$          & $^oCm^2/kWh$   & 0.3415               & 0.1237               & 0.1177               \\ \hline
		$\tau_{wz,f}$      & hours          & 18.7779              & 68.4388              & 100                  \\ \hline
		$\tau_{wa,f}$      & hours          & 4157.5               & 1145.7               & 1129.2               \\ \hline
		$A_{w,f}$          & $^oCm^2/kWh$   & 9.9$\times 10^{-5}$ & 3.42$\times 10^{-4}$ & 3.75$\times 10^{-4}$ \\ \hline
	\end{tabular}
\end{table}
\egroup
\begin{figure}[t]
	\centering
	\includegraphics[width=0.99\linewidth]{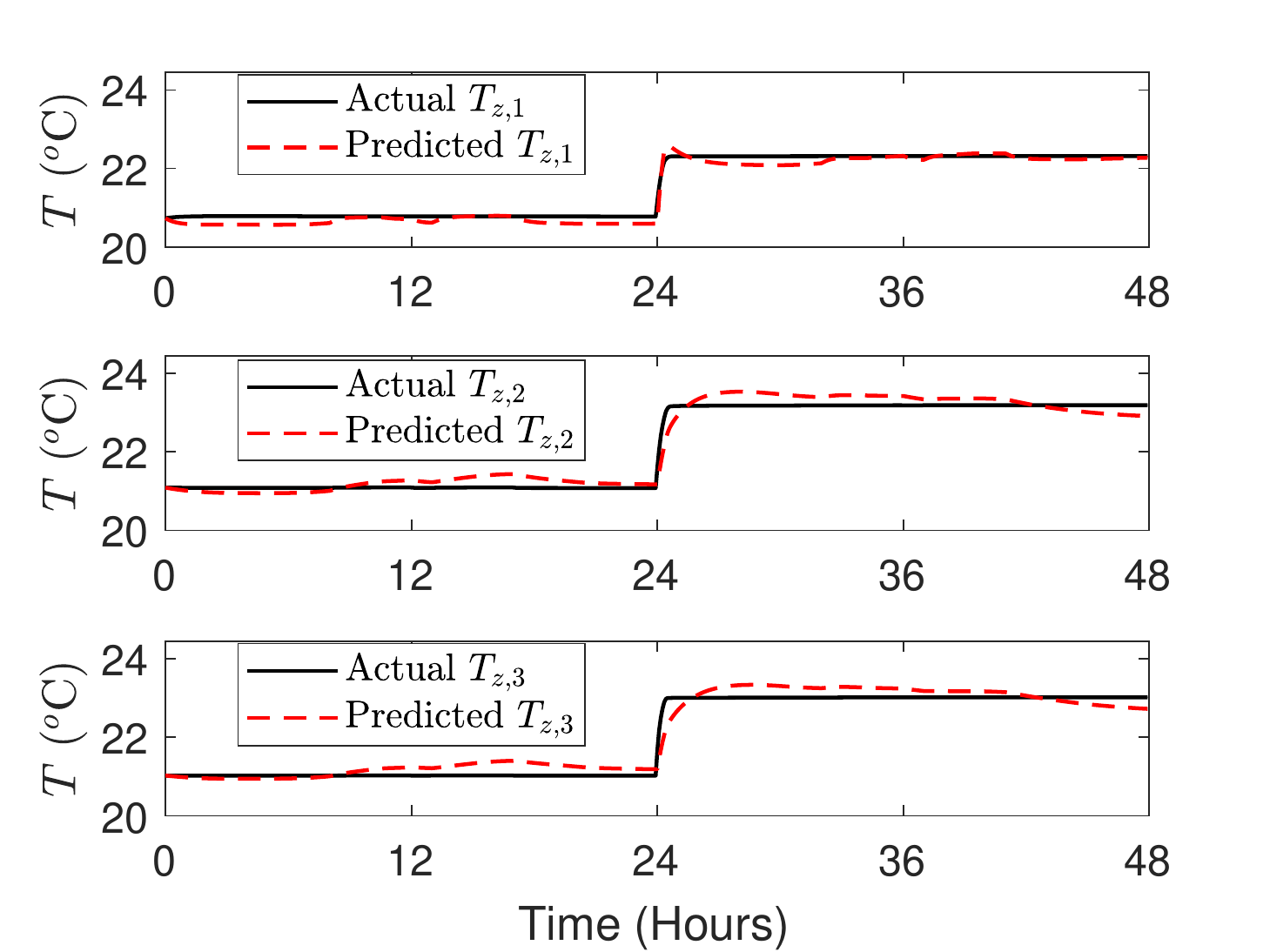}
	\caption{Out of sample aggregate zone temperature ($T_{z,f}$) prediction results using the estimated aggregate RC network model.}
	\label{fig:out_of_sample_thermal_model_for_mpc_si}\vspace{-10pt}
      \end{figure}
      \fi
\begin{figure}[b]
	\centering
	\includegraphics[width=0.99\linewidth]{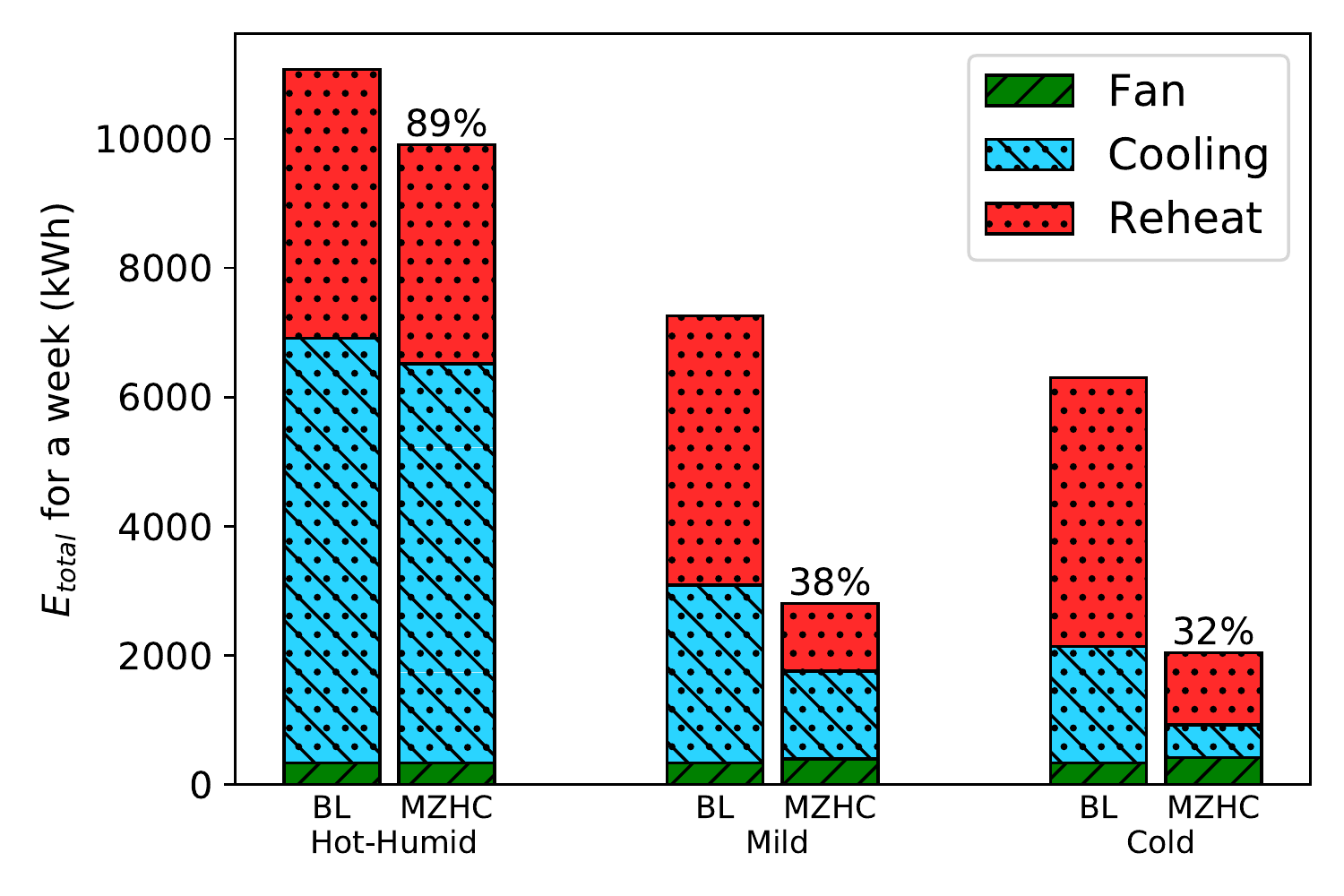}
	\caption{Comparison of the total energy consumed for a week when using the baseline (\bl) and proposed (\slmpc) controllers for different outdoor weather conditions.}
	\label{fig:energy_consumption}\vspace{-10pt}
\end{figure}
\begin{figure*}[!t]
	\centering
	\subfigure[Outdoor weather data used in simulations (outdoor air temperature, outdoor air relative humidity, and solar irradiance).\label{fig:outdoor_weather_hot}]{\includegraphics[width=0.485\textwidth]{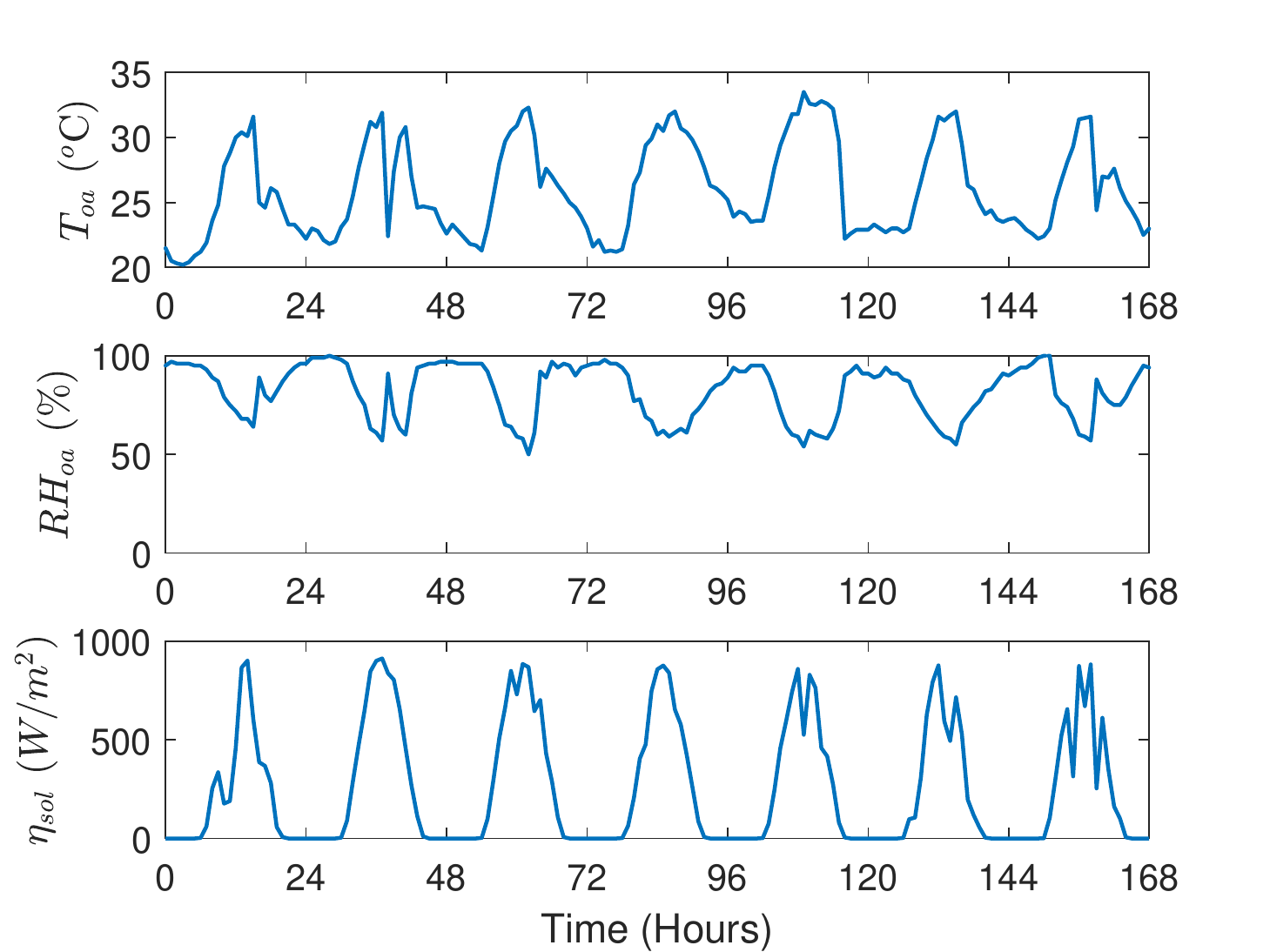}}\quad
	\subfigure[Comparing the power consumptions (fan, cooling, and total reheat power).\label{fig:comparison_power_hot}]{\includegraphics[width=0.485\textwidth]{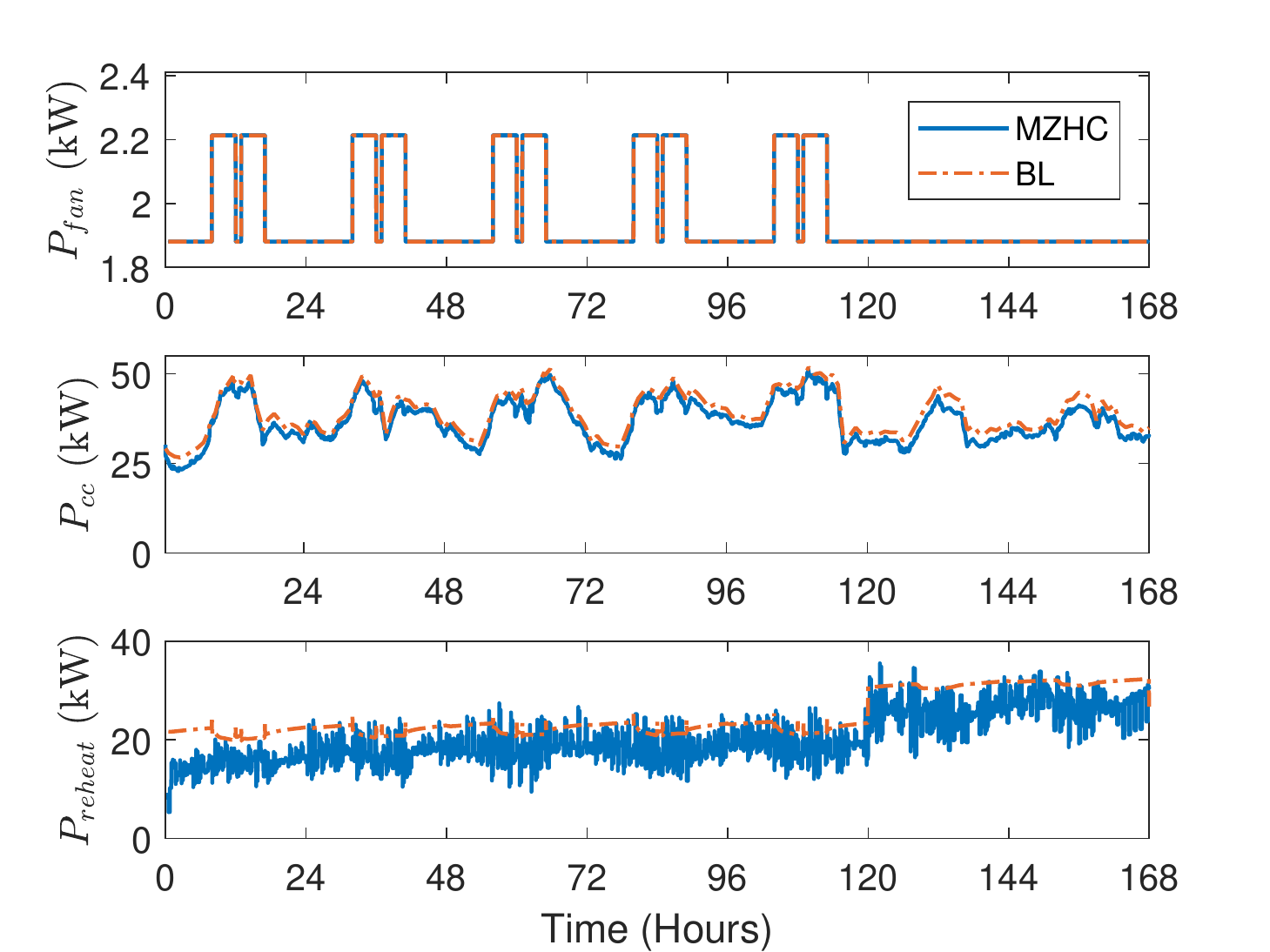}}
	\subfigure[Aggregate conditions of floors/\metazones{} 1, 2, and 3 (temperatures and relative humidities) when using \slmpc{} and \bl{}. The black dashed lines are the thermal comfort limits.\label{fig:aggregate_conditions_hot}]{\includegraphics[width=0.485\textwidth]{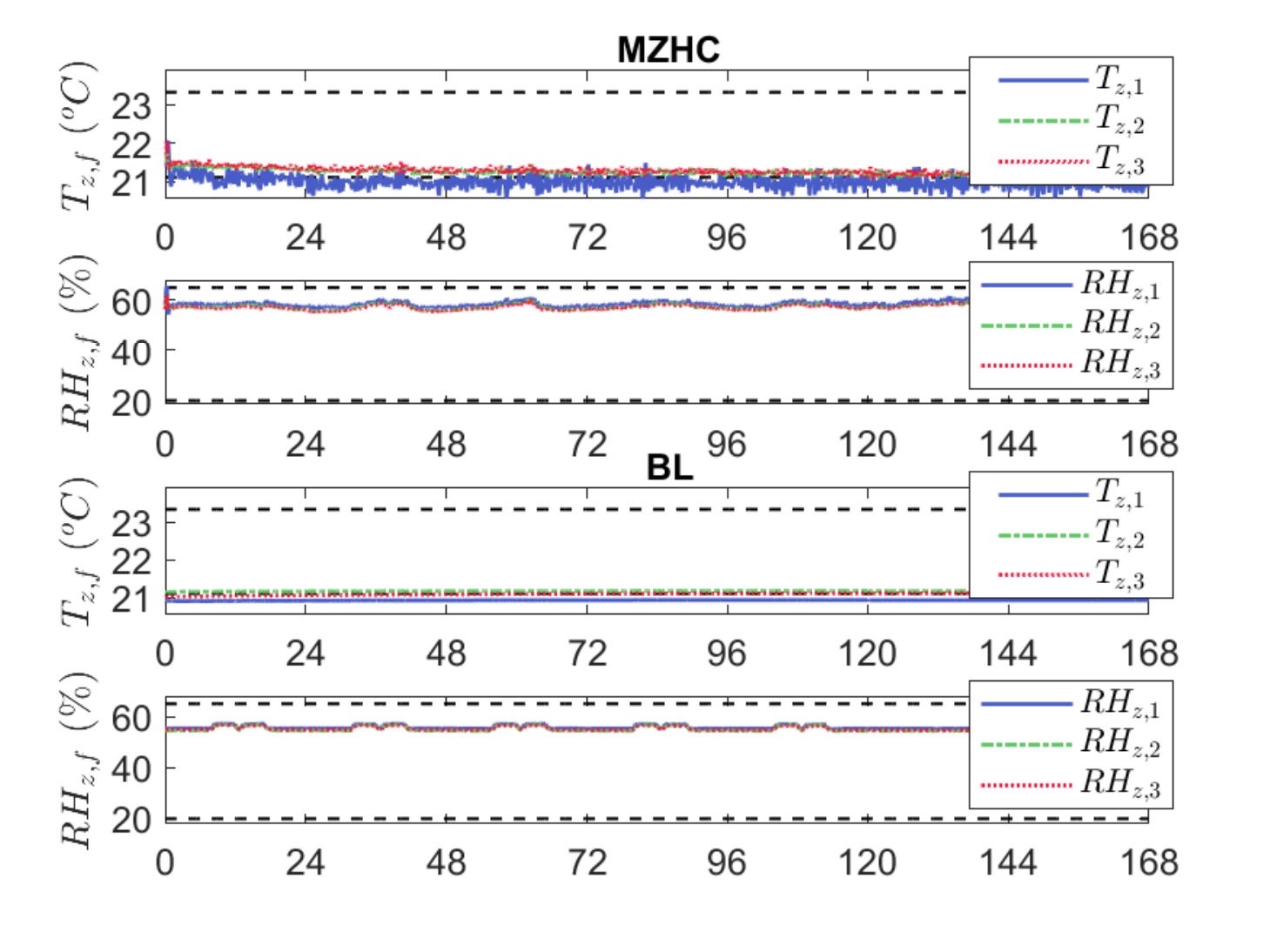}}\quad
	\subfigure[Conditions at the AHU (supply airflow rate, outdoor airflow rate, conditioned air temperature, and conditioned air humidity ratio).\label{fig:control_commands_hot}]{\includegraphics[width=0.485\textwidth]{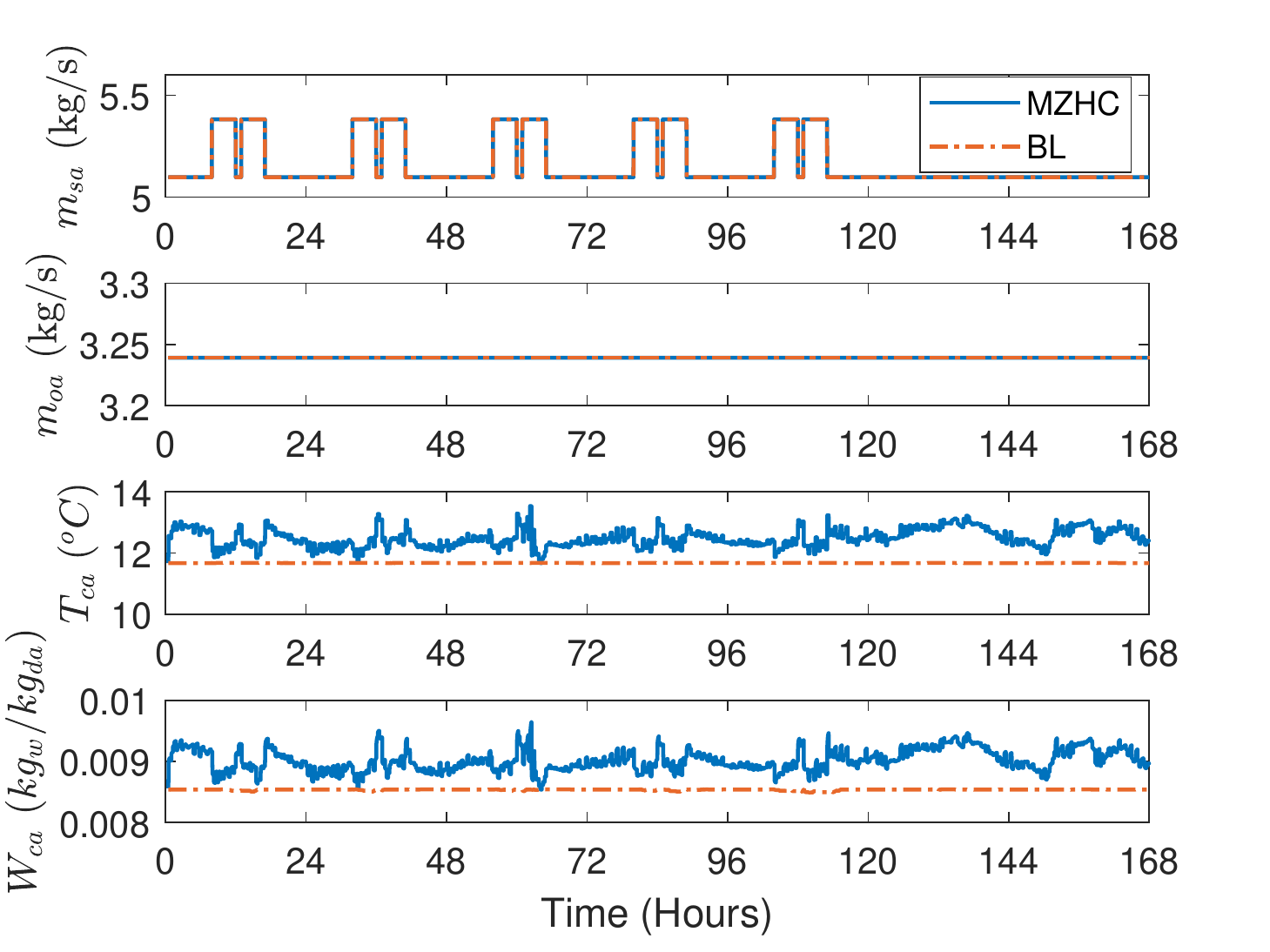}}
	\caption{Comparison of the two controllers for a hot-humid week (Jul/06 to Jul/13, Gainesville, Florida, USA).} \label{fig:mpc_simulation_results_hot}\vspace{-10pt}
\end{figure*}
\begin{figure}[!tb]
	\centering
	\includegraphics[width=0.99\linewidth]{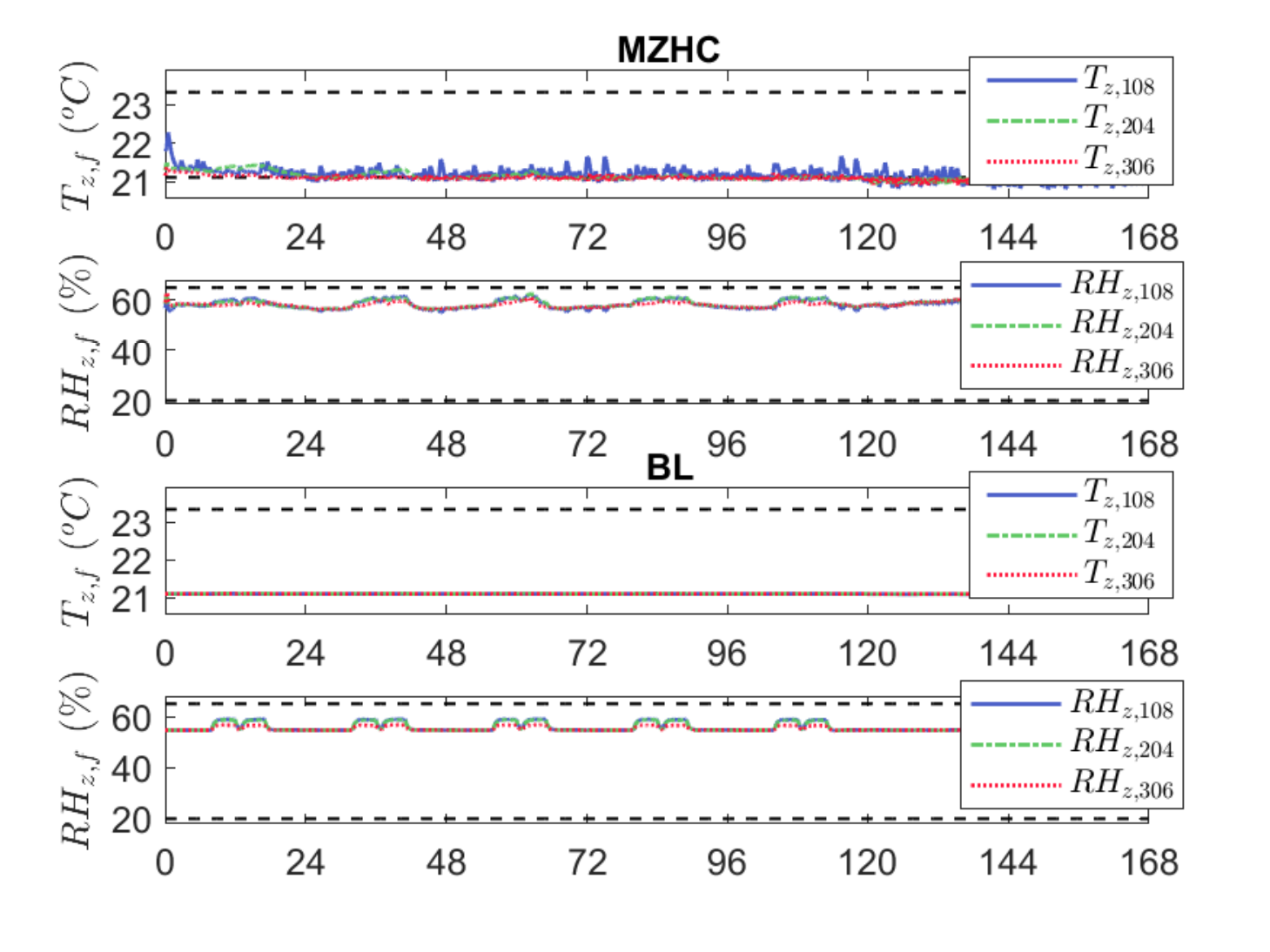}\vspace{-15pt}
	\caption{Individual zone conditions (temperatures and relative humidities) when using \slmpc{} and \bl{} for a hot-humid week. The black dashed lines are the thermal comfort limits.}
	\label{fig:individual_zone_conditions_hot}\vspace{-10pt}
\end{figure}
\ifArxivVersion
\begin{figure*}[!t]
	\centering
	\subfigure[Outdoor weather data used in simulations (outdoor air temperature, outdoor air relative humidity, and solar irradiance).\label{fig:outdoor_weather_mild}]{\includegraphics[width=0.485\textwidth]{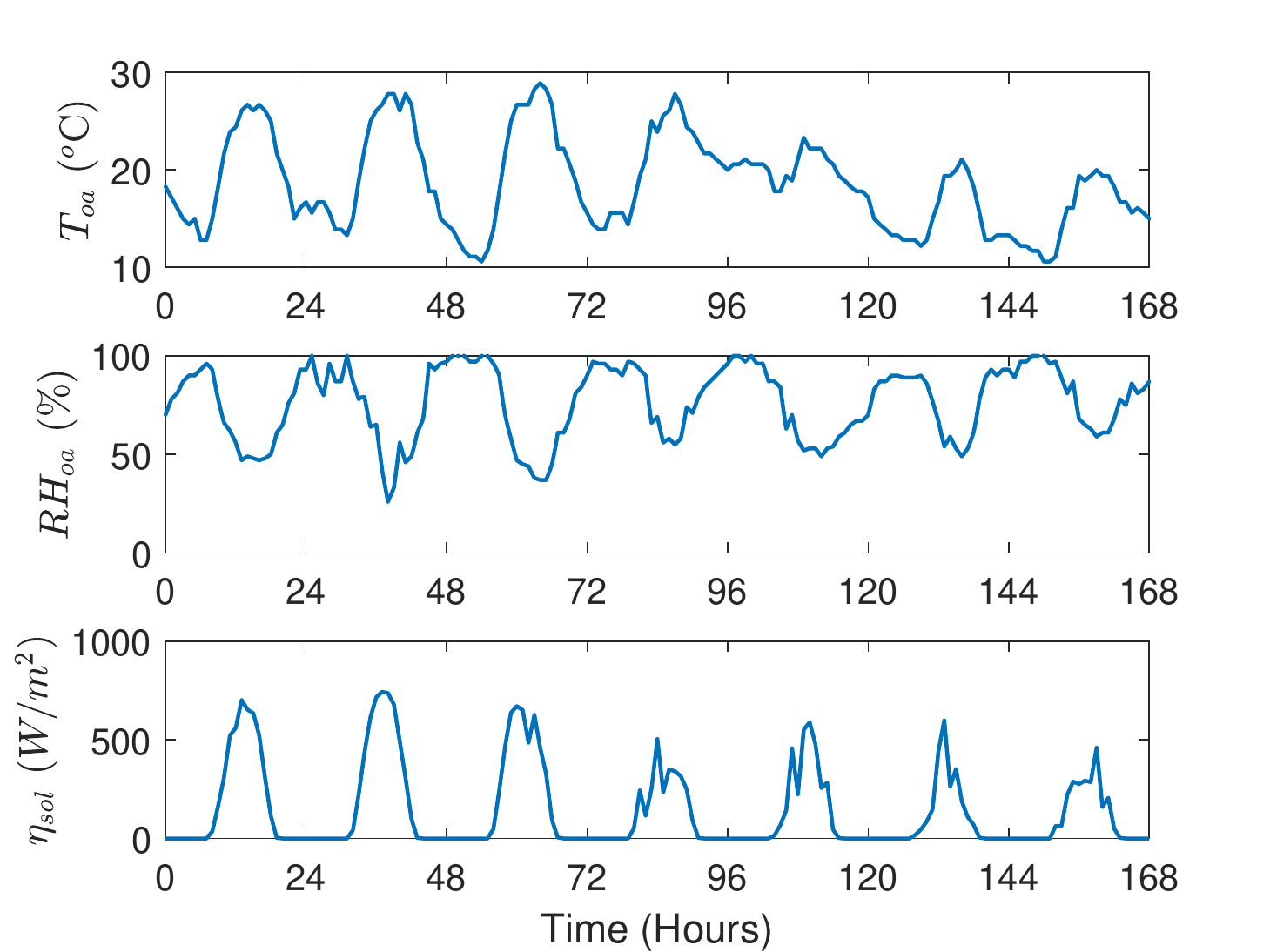}}\quad
	\subfigure[Comparing the power consumptions (fan, cooling, and total reheat power).\label{fig:comparison_power_mild}]{\includegraphics[width=0.485\textwidth]{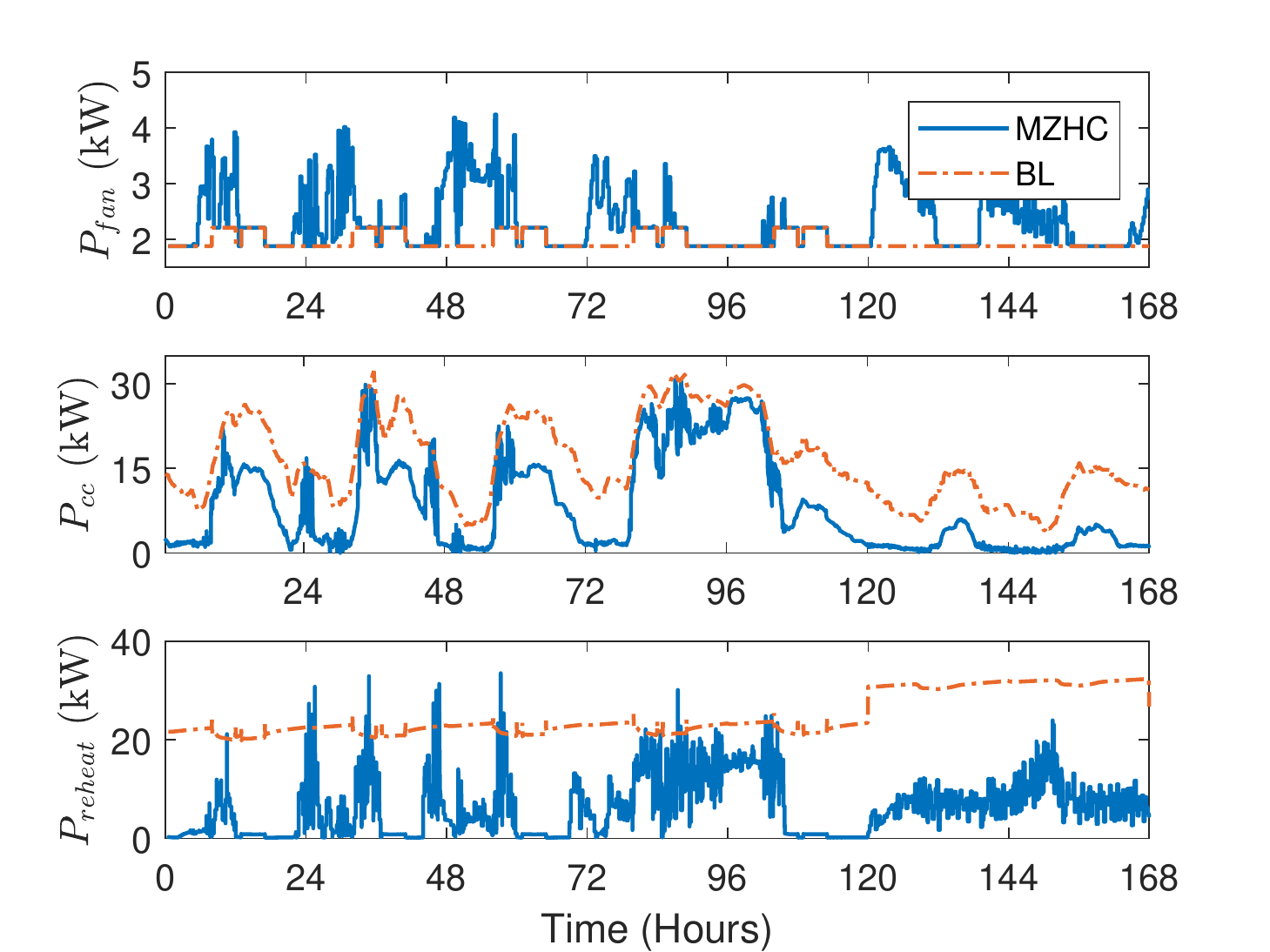}}
	\subfigure[Aggregate conditions of floors/\metazones{} 1, 2, and 3 (temperatures and relative humidities) when using \slmpc{} and \bl{}. The black dashed lines are the thermal comfort limits.\label{fig:aggregate_conditions_mild}]{\includegraphics[width=0.485\textwidth]{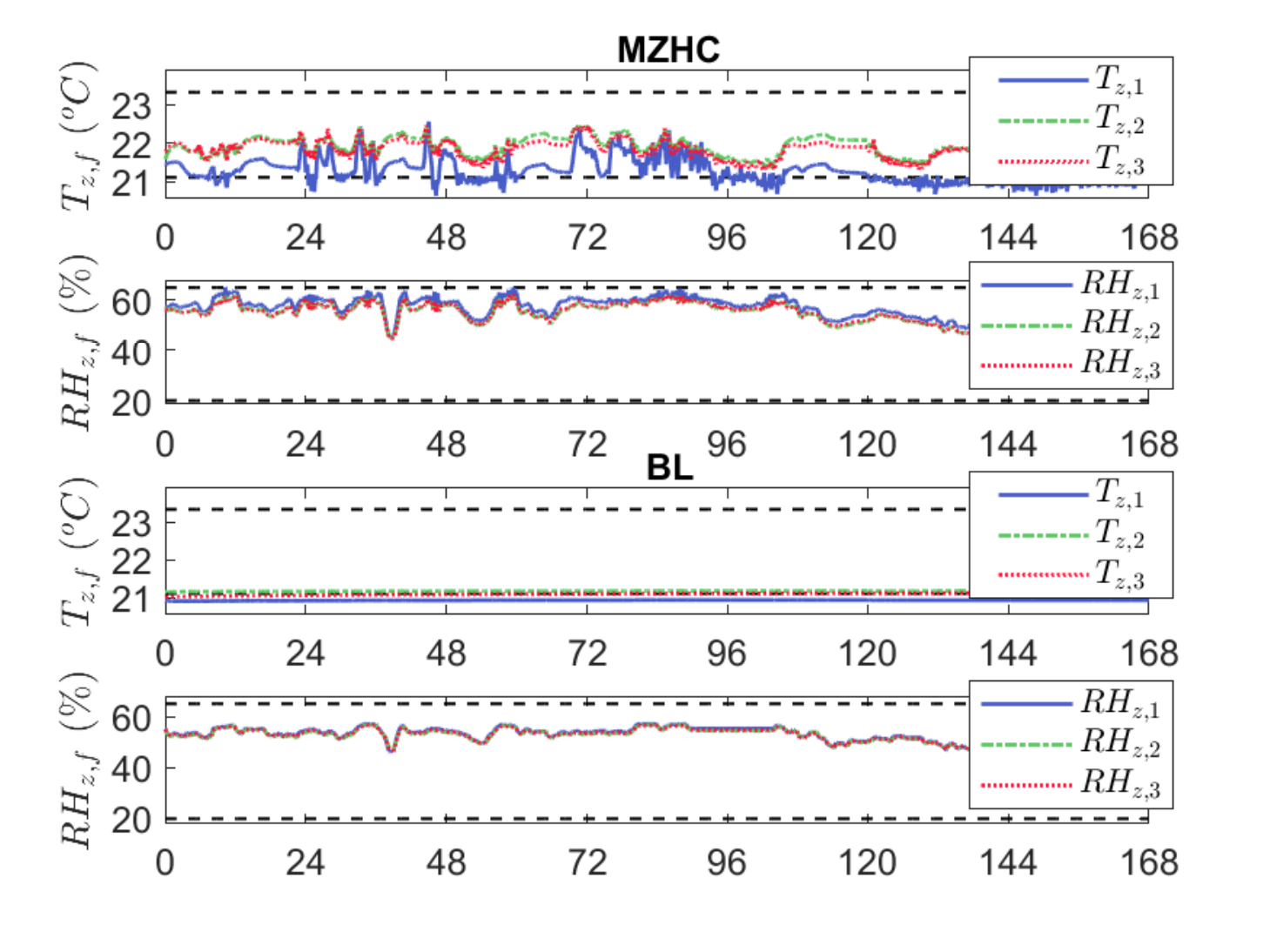}}\quad
	\subfigure[Conditions at the AHU (supply airflow rate, outdoor airflow rate, conditioned air temperature, and conditioned air humidity ratio).\label{fig:control_commands_mild}]{\includegraphics[width=0.485\textwidth]{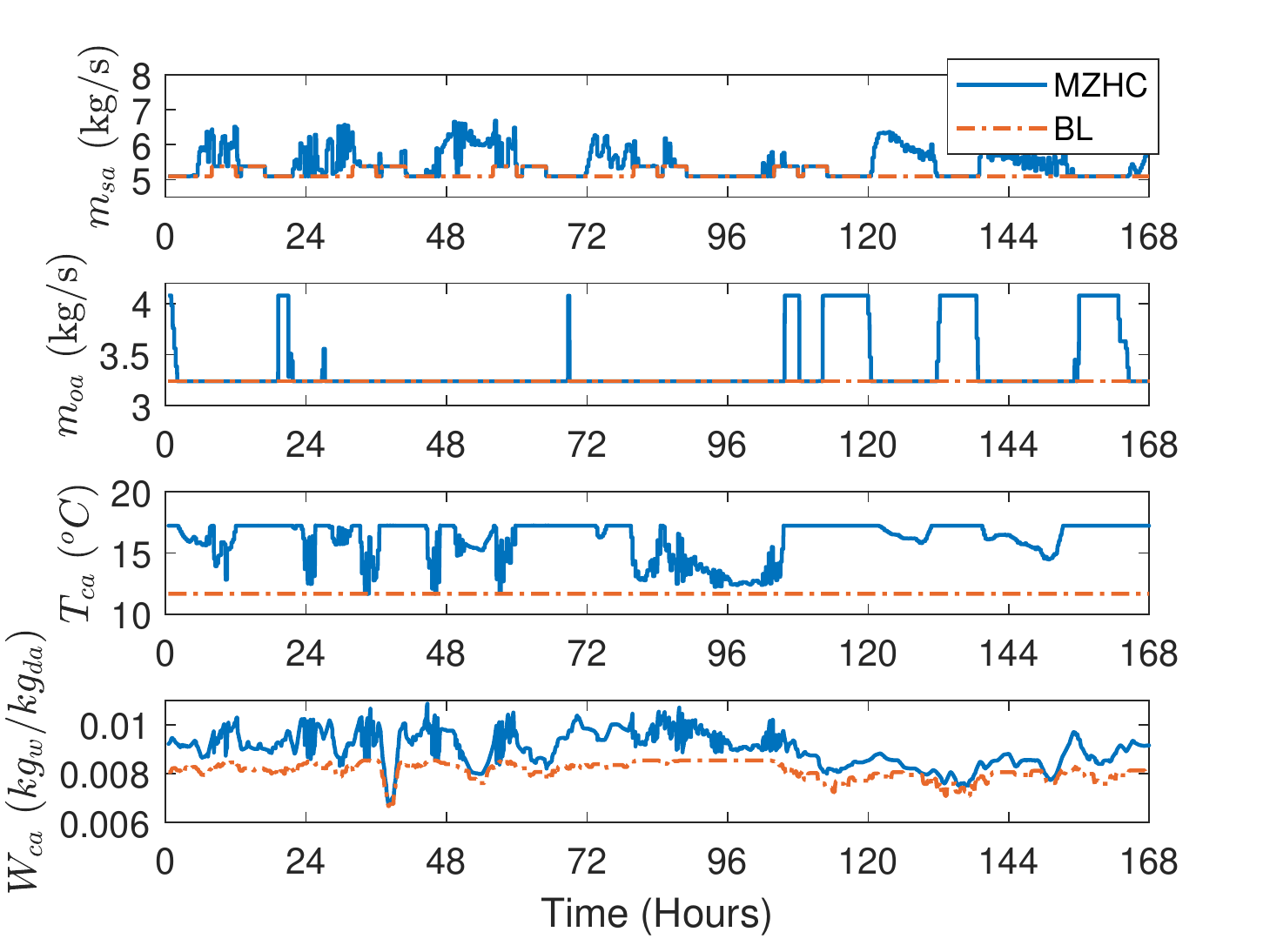}}
	\caption{Comparison of the two controllers for a mild week (Feb/19 to Feb/26, Gainesville, Florida, USA).} \label{fig:mpc_simulation_results_mild}\vspace{-10pt}
\end{figure*}
\fi

\textbf{MZHC Parameters:} The optimization problems in \hlc{} and \llc{} are solved using CasADi~\cite{Andersson2019} and IPOPT~\cite{wacbie:2006}, a nonlinear programming (NLP) solver, on a Desktop Windows computer with 16GB RAM and a 3.60 GHZ $\times$ 8 CPU. As mentioned in Section~\ref{sec:HLC}, $\Delta t=5$~minutes, $\Delta T=15$~minutes, $T=24$~hours, $N=96$, and $M=3$. The number of decision variables for the \hlc{} are 7008 and for the \llc{} are 54. On an average it takes only 3.28~seconds to solve the optimization problem in the \hlc{} and 0.018~seconds to solve the optimization problem in the \llc{}. 

The parameters for the control-oriented cooling and dehumidifying coil model are fit using the procedure explained in~\cite{RamanMPC_AE:2020}. For the validation data set, the root mean square errors are 0.97$\degree$C (1.75$\degree$F, 7.6\%) for $T_{ca}$ and 0.63$\times10^{-4}kg_w/kg_{da}$ (7.6\%) for $W_{ca}$. 

Since the Innovation Hub building has three floors, we aggregate it into three meta-zones, i.e., $f\in\{1,2,3\}\defeq\mathbf{F}$. The parameters of the aggregate thermal dynamics model for each meta-zone are estimated using the algorithm presented in~\cite{GuoAggregationEnB:2020}. \ifArxivVersion The parameters are shown in Table~\ref{table:thermal_model_mpc}. Figure~\ref{fig:out_of_sample_thermal_model_for_mpc_si} shows the out of sample prediction results using the estimated aggregate RC network model. \else The interested reader is referred to~\cite{RamanMPC-Based_AriXiV:2021} for the values of the parameters and out-of-sample prediction accuracy of the identified model. \fi

For the aggregate humidity dynamics model, floor volumes used are $V_1=1036.6~m^3$, $V_2=1504.1~m^3$, and $V_3=1330.8~m^3$, which are obtained from mechanical drawings of the building.

The following limits are used for the zone temperature constraint~\eqref{eq:hlc_T_z_box}: $T_{z,f}^{low}=$21.1$\degree$C (70$\degree$F) and $T_{z,f}^{high}=$23.3$\degree$C (74$\degree$F). The coefficients for the humidity constraint in~\eqref{eq:hlc_W_z_box} are $a^{high}= 0.000621~kg_w/kg_{da}/\degree C$ and $b^{high}= -0.173323~kg_w/kg_{da}$, which corresponds to a relative humidity of 60\%, and $a^{low}= 0.000203~kg_w/kg_{da}/\degree C$ and $b^{low}= -0.056516~kg_w/kg_{da}$, which corresponds to a relative humidity of 20\%. We introduce a factor of safety by using a slightly tighter higher limit of 60\% for the relative humidity of the zones when compared to the thermal comfort envelope presented in Figure~\ref{fig:thermal_comfort_envelope}.

The minimum allowed value for the outdoor airflow rate~($m_{oa}^{min}$) is 3.24~kg/s~(5700~cfm), which is obtained from the AHU schedule in the mechanical drawings for the building.  The maximum possible value for the outdoor airflow rate~($m_{oa}^{max}$) is 8.52~kg/s~(15000~cfm). The various limits on the supply airflow rates are obtained using the VAV schedule \ifArxivVersion presented in Table~\ref{table:vav_schedule}. \else for the building. \fi The remaining limits used in the controllers are as follows: $r_{oa}^{low}=0\%$, $r_{oa}^{high}=100\%$, $\tilde T_z^{db}=0.56\degree$C (1$\degree$F), $T_{ca}^{low}=11.67\degree$C (53$\degree$F), $T_{ca}^{high}=17.2\degree$C (63$\degree$F), and $T_{sa}^{high}=30\degree$C (86$\degree$F). The higher limit on the conditioned air temperature ($T_{ca}^{high}$) is to introduce a factor of safety and make the controller robust.

The MPC controller requires predictions of the various exogenous inputs specified in~\eqref{eq:hlc_exogenous_inputs}. We compute the loads due to occupants in $q_{int,f}$ and $\omega_{int,f}$ based on the occupancy profile used in simulating the plant. The outdoor weather related exogenous inputs are assumed to be fully known.

\textbf{BL Parameters:} The cooling and heating setpoints are chosen to be 21.1$\degree$C (70$\degree$F) and 23.3$\degree$C (74$\degree$F), respectively. The minimum, maximum, and heating maximum values for the supply airflow rate of all the VAV boxes are \ifArxivVersion listed in Table~\ref{table:vav_schedule}. \else obtained from the VAV schedule for the building. \fi The maximum allowed value for the supply air temperature ($T_{sa}^{high}$) is 30$\degree$C (86$\degree$F). The conditioned air temperature ($T_{ca}$) is kept at a constant value of 11.67$\degree$C (53$\degree$F). Typically $T_{ca}$ is kept at 12.8$\degree$C (55$\degree$F), especially in hot-humid climates, to ensure humidity control~\cite{WilliamsWhy:2013}, but \ifArxivVersion recall that \fi we assume there is a 1.11$\degree$C (2$\degree$F) increase in temperature because of the heat from the draw-through supply fan in the AHU, so we keep it at 11.67$\degree$C (53$\degree$F) to compensate for it. The outdoor airflow rate is kept at 3.24~kg/s~(5700~cfm), which is obtained from mechanical drawings for the building.

\section{Results and Discussions}\label{sec:results}
Performance of the controllers is compared using three types of outdoor weather conditions: hot-humid~(Jul/06 to Jul/13), mild~(Feb/19 to Feb/26), and cold (Jan/30 to Feb/06). The proposed controller reduces energy use significantly compared to \bl{}, from approximately 11\% to 68\% depending on weather; see Figure~\ref{fig:energy_consumption}. The indoor climate control performance of  \slmpc{}  and \bl{} are nearly identical. With \slmpc{}, the RMSE (root mean square error) of zone temperature violation is 0.1$\degree$C (0.18$\degree$F) and the RMSE of zone relative humidity violation is 0.05\%, while with \bl{} they are 0.01$\degree$C (0.02$\degree$F) and 0\% respectively. The computational cost of the proposed \slmpc{} is small, just a few seconds are needed to compute decisions at every control update. On an average it takes 3.28~seconds to solve the optimization problem in the \hlc{} and 0.018~seconds to solve the optimization problem in the \llc{}.

Simulation results for the different weather conditions are discussed in detail next.

\subsection{Hot-Humid Week}\label{sec:hot_humid}
Figure~\ref{fig:mpc_simulation_results_hot} shows the simulation time traces for a hot-humid week. It is found that using the proposed \slmpc{} leads to 11\% energy savings when compared to \bl{}, as presented in Figure~\ref{fig:energy_consumption}.

Both controllers lead to negligible violations of aggregate zone temperature ($T_{z,f}$) and relative humidity ($RH_{z,f}$) constraints. Data for the three \metazones{} are shown in Figure~\ref{fig:aggregate_conditions_hot}, and for three individual zones, one from each floor, are shown in Figure~\ref{fig:individual_zone_conditions_hot}. \bl{} ensures that dry enough air is supplied to the zones at all times by keeping the conditioned air temperature ($T_{ca}$) at a constant low value of 11.67$\degree$C (53$\degree$F), and hence the humidity limit is not violated. In the case of \slmpc{}, the humidity constraint is found to be active always. This can be seen in Figure~\ref{fig:aggregate_conditions_hot}; recall that we use a tighter constraint of 60\% instead of 65\% to introduce a factor of safety. This active constraint limits the increase in $T_{ca}$, which can be seen in Figure~\ref{fig:control_commands_hot}. One of the reasons reheating is required even during such a hot week (the outdoor air temperature is as high as 32.2$\degree$C/90$\degree$F) is because of this active humidity constraint, which requires dry, and thus, cold air to be supplied to the zones. This could also be one of the reasons \slmpc{} decides to maintain the zone temperatures at the lower limit (see Figure~\ref{fig:aggregate_conditions_hot}). Keeping the zones at a higher temperature will require an increase in reheating energy. 

Most of the prior works on using MPC for HVAC control ignore humidity and latent heat in their formulation. In an attempt to reduce energy/cost, such controllers are likely to make decisions during these hot-humid conditions which will lead to poor indoor humidity, as they are unaware of the factors mentioned above~\cite{RamanMPC_AE:2020}. Thus such controllers cannot be used particularly in hot-humid climates.

The energy savings by \slmpc{} is because of two main reasons. One, \slmpc{} increases the $T_{ca}$ as long as the humidity constraints are not violated, while \bl{} uses a conservatively designed value as explained above. This leads to a reduction in cooling energy consumption by \slmpc{}; see Figures~\ref{fig:energy_consumption} and \ref{fig:comparison_power_hot}. Two, the warmer $T_{ca}$ supplied to the VAV boxes requires lesser reheating in the case of \slmpc{}. This leads to a reduction in the reheat energy consumption, which can be seen in Figures~\ref{fig:energy_consumption} and \ref{fig:comparison_power_hot}. The decisions regarding the supply airflow rates are found to be the same for both the controllers, and thus the fan energy consumptions are identical. 

Since the outdoor air is hot and humid (see Figure~\ref{fig:outdoor_weather_hot}), bringing in more than the minimum outdoor airflow rate ($m_{oa}$) required will increase the sensible and latent loads on the cooling coil. So \slmpc{} decides to keep $m_{oa}$ at the minimum; see Figure~\ref{fig:control_commands_hot}.

\ifArxivVersion
\subsection{Mild Week}\label{sec:mild}

Figure~\ref{fig:mpc_simulation_results_mild} shows the simulation results for a mild week. It is found that using \slmpc{} leads to $\sim$60\% energy savings when compared to \bl{}, as shown in Figure~\ref{fig:energy_consumption}. This significant reduction in energy consumption can be attributed to three main reasons. Two of them are the same as those explained in Section~\ref{sec:hot_humid}; the effects here are more prominent and the details are discussed in the subsequent paragraph. The third is that, since the outdoor weather is mild and dry (see Figure~\ref{fig:outdoor_weather_mild}), \slmpc{} also decides to use ``free'' cooling when possible by bringing in more than the minimum outdoor air required which leads to further reduction in cooling energy consumption.
 
Figure~\ref{fig:aggregate_conditions_mild} shows the aggregate zone temperature ($T_{z,f}$) and relative humidity ($RH_{z,f}$) for all the three floors/\metazones{}. Unlike the results for the hot-humid week, the humidity constraint is found to be only intermittently active as the outdoor weather is relatively dry. This provides more room for optimizing the conditioned air temperature, which has two important implications: (i)~reduction in the cooling energy consumption, and (ii)~minimal reheat energy consumption. For example, see 13:00-22:00~h in Figure~\ref{fig:control_commands_mild} where $T_{ca}$ is at its higher limit, and also see Figure~\ref{fig:comparison_power_mild} where $P_{cc}$ is significantly reduced and $P_{reheat}$ is almost zero.

\subsection{Cold Week}\label{sec:cold}
The energy savings when using \slmpc{} is found to be significant as can be seen in Figure~\ref{fig:energy_consumption}. Since the outdoor weather is cold and dry, there is a lot of room for optimizing the control commands. The reasons for energy reduction when using \slmpc{} are the same as those explained in Section~\ref{sec:mild}, therefore we do not discuss them in detail here.

\else
\subsection{Mild and Cold Weeks}\label{sec:cold}
As seen from Figure~\ref{fig:energy_consumption}, energy savings are higher than 60\% during mild and cold weeks. This significant reduction in energy consumption can be attributed to three main reasons. Two of them are the same as those explained in Section~\ref{sec:hot_humid}; but the effects here are more prominent. For instance, unlike the hot-humid week, the humidity constraint is only intermittently active since the outdoor weather is drier. This provides more room for optimizing the conditioned air temperature, which has two important implications: (i)~reduction in the cooling energy consumption, and (ii)~minimal reheat energy consumption. The third is that, since the outdoor weather is cold and dry, \slmpc{} also decides to use ``free'' cooling when possible by bringing in more than the minimum outdoor air required which leads to further reduction in cooling energy consumption.

We omit the detailed results for the mild and cold weeks here; the interested reader can find them in~\cite{RamanMPC-Based_AriXiV:2021}.
\fi

\begin{comment}\label{com:single_maximum}
Recall that as mentioned in Section~\ref{sec:baseline}, the Innovation Hub building uses \emph{Single Maximum} algorithm for zone climate control as opposed to the \emph{Dual Maximum} algorithm presented here~\cite{ASHRAE_handbook_applications:11}. In the case of \emph{Single Maximum} algorithm, the minimum allowed value for the supply airflow rate has to be high enough so that the heating load can be met with low enough supply air temperature to prevent thermal stratification~\cite{ASHRAE_handbook_applications:11}. While in the case of \emph{Dual Maximum} algorithm, the airflow rate is varied during the heating mode, thus the minimum allowed airflow rate is not limited by stratification. Therefore using \emph{Single Maximum} algorithm leads to higher fan, cooling, and reheating energy consumptions. This also implies that the energy savings by the proposed controller will be even higher.
\end{comment}

\section{Conclusion}\label{sec:conclusions}
The proposed control architecture is designed to address a number of limitations in the existing literature on multi-zone building control using MPC. The main one is the reliance on a \hiresmodel{} multi-zone model, which can be challenging to obtain. 
A \loresmodel{} model of the building is more convenient since such a model can be identified in a tractable manner from measurements. The challenge then is to convert the MPC decisions, which are computed for the fictitious zones in the model, to the commands for the VAV boxes of the actual building. The proposed architecture does that by posing this conversion as a projection problem that uses not only what the MPC computes but also feedback from zones. The result is a principled method of computing VAV box commands that are consistent with the optimal decisions made by the MPC without needing dynamic models of indiviudal zones. At the same time, the use of feedback from the zones ensures that zone climate states are also close to the aggregate climate states computed by the MPC.  

The positive simulation results provide confidence on the effectiveness of the proposed controller especially because of the large plant model mismatch. The simulation testbed mimics a real building closely, including the heterogenous nature of the zones in the building. 
Another observation from the simulations that should be emphasized is the need to incorporate humidity and latent heat. The indoor humidity constraint is seen to be active when using the proposed controller, especially during hot-humid weather. Without humidity being explicitly considered, the controller is likely to have caused high space humidity in an effort to reduce energy use.

There are many avenues for further exploration, such as experimental verification, modifying the formulation to minimize cost instead of energy and to include demand charges, improving methods to forecast disturbances, etc.

%%%%%%%%%%%%%%%%%%%%%%%%%%%%%%%%%%%%%%%%%%%%%%%%%%%%%%%%%%%%%%%%%%%%%%
\begin{acknowledgment}
This research reported here has been partially supported by the NSF through award \# 1934322 and the State of Florida through a REET (Renewable Energy and Energy Efficient Technologies) grant. We thank David Brooks for help in understanding the design and operation of Innovation Hub's HVAC system.  
\end{acknowledgment}

%%%%%%%%%%%%%%%%%%%%%%%%%%%%%%%%%%%%%%%%%%%%%%%%%%%%%%%%%%%%%%%%%%%%%%
% The bibliography is stored in an external database file
% in the BibTeX format (file_name.bib).  The bibliography is
% created by the following command and it will appear in this
% position in the document. You may, of course, create your
% own bibliography by using thebibliography environment as in
%
% \begin{thebibliography}{12}
% ...
% \bibitem{itemreference} D. E. Knudsen.
% {\em 1966 World Bnus Almanac.}
% {Permafrost Press, Novosibirsk.}
% ...
% \end{thebibliography}

% Here's where you specify the bibliography style file.
% The full file name for the bibliography style file 
% used for an ASME paper is asmems4.bst.
\bibliographystyle{asmems4}

% Here's where you specify the bibliography database file.
% The full file name of the bibliography database for this
% article is asme2e.bib. The name for your database is up
% to you.
%\bibliography{asme2e}
\bibliography{\DiCEbibPATH/Barooah,\DiCEbibPATH/optimization,\DiCEbibPATH/systemid,\DiCEbibPATH/grid,\DiCEbibPATH/building,\DiCEbibPATH/ControlTheory,\DiCEbibPATH/Building_Parameter_Est,\NRbibPATH/Raman}

\begin{thebibliography}{10}

\bibitem{serale2018model}
Serale, G., Fiorentini, M., Capozzoli, A., Bernardini, D., and Bemporad, A.,
  2018.
\newblock ``Model predictive control {(MPC)} for enhancing building and {HVAC}
  system energy efficiency: {P}roblem formulation, applications and
  opportunities''.
\newblock {\em Energies, {\bf 11}}(3), p.~631.

\bibitem{shaikh2014review}
Shaikh, P.~H., Nor, N. B.~M., Nallagownden, P., Elamvazuthi, I., and Ibrahim,
  T., 2014.
\newblock ``A review on optimized control systems for building energy and
  comfort management of smart sustainable buildings''.
\newblock {\em Renewable and Sustainable Energy Reviews, {\bf 34}}, pp.~409 --
  429.

\bibitem{goybar:CDC:2013}
Goyal, S., and Barooah, P., 2013.
\newblock ``Energy-efficient control of an air handling unit for a single-zone
  {VAV} system''.
\newblock In {IEEE} Conference on Decision and Control, pp.~4796 -- 4801.

\bibitem{JoeModel:2019}
Joe, J., and Karava, P., 2019.
\newblock ``A model predictive control strategy to optimize the performance of
  radiant floor heating and cooling systems in office buildings''.
\newblock {\em Applied Energy, {\bf 245}}, pp.~65 -- 77.

\bibitem{RamanMPC_AE:2020}
Raman, N., Devaprasad, K., Chen, B., Ingley, H.~A., and Barooah, P., 2020.
\newblock ``Model predictive control for energy-efficient {HVAC} operation with
  humidity and latent heat considerations''.
\newblock {\em Applied Energy, {\bf 279}}, December, p.~115765.

\bibitem{ChenOccupant:2016}
Chen, X., Wang, Q., and Srebric, J., 2016.
\newblock ``Occupant feedback based model predictive control for thermal
  comfort and energy optimization: A chamber experimental evaluation''.
\newblock {\em Applied Energy, {\bf 164}}, pp.~341 -- 351.

\bibitem{ma2012demand}
Ma, J., Qin, J., Salsbury, T., and Xu, P., 2012.
\newblock ``Demand reduction in building energy systems based on economic model
  predictive control''.
\newblock {\em Chemical Engineering Science, {\bf 67}}(1), pp.~92 -- 100.
\newblock Dynamics, Control and Optimization of Energy Systems.

\bibitem{Bengeaetal:2013}
Bengea, S.~C., Kelman, A.~D., Borrelli, F., Taylor, R., and Narayanan, S.,
  2013.
\newblock ``Implementation of model predictive control for an {HVAC} system in
  a mid-size commercial building''.
\newblock {\em {HVAC\&R Research}, {\bf 20}}, pp.~121--135.

\bibitem{RadhakrishnanToken:2016}
Radhakrishnan, N., Su, Y., Su, R., and Poolla, K., 2016.
\newblock ``Token based scheduling for energy management in building {HVAC}
  systems''.
\newblock {\em Applied Energy, {\bf 173}}, pp.~67 -- 79.

\bibitem{PngInternet:2019}
Png, E., Srinivasan, S., Bekiroglu, K., Chaoyang, J., Su, R., and Poolla, K.,
  2019.
\newblock ``An internet of things upgrade for smart and scalable heating,
  ventilation and air-conditioning control in commercial buildings''.
\newblock {\em Applied Energy, {\bf 239}}, pp.~408 -- 424.

\bibitem{ma2012distributed}
Ma, Y., Richter, S., and Borrelli, F., 2012.
\newblock ``Chapter 14: Distributed model predictive control for building
  temperature regulation''.
\newblock {\em {In} Control and Optimization with Differential-Algebraic
  Constraints, {\bf 22}}, March, pp.~293--314.

\bibitem{mei2018multizone}
{Mei}, J., and {Xia}, X., 2018.
\newblock ``Multi-zone building temperature control and energy efficiency using
  autonomous hierarchical control strategy''.
\newblock In 2018 IEEE 14th International Conference on Control and Automation
  (ICCA), pp.~884--889.

\bibitem{patel2016distributed}
{Patel}, N.~R., {Risbeck}, M.~J., {Rawlings}, J.~B., {Wenzel}, M.~J., and
  {Turney}, R.~D., 2016.
\newblock ``Distributed economic model predictive control for large-scale
  building temperature regulation''.
\newblock In 2016 American Control Conference (ACC), pp.~895--900.

\bibitem{yang2020hvac}
{Yang}, Y., {Hu}, G., and {Spanos}, C.~J., 2020.
\newblock ``{HVAC} energy cost optimization for a multizone building via a
  decentralized approach''.
\newblock {\em IEEE Transactions on Automation Science and Engineering, {\bf
  17}}(4), pp.~1950--1960.

\bibitem{long2016hierarchical}
{Yushen Long}, {Shuai Liu}, {Lihua Xie}, and {Johansson}, K.~H., 2016.
\newblock ``A hierarchical distributed {MPC} for {HVAC} systems''.
\newblock In 2016 American Control Conference (ACC), pp.~2385--2390.

\bibitem{jorissen2016towards}
Jorissen, F., and Helsen, L., 2016.
\newblock ``Towards an automated tool chain for {MPC} in multi-zone
  buildings''.
\newblock In 4th International Conference on High Performance Buildings,
  pp.~1--10.

\bibitem{liwen:2014}
Li, X., and Wen, J., 2014.
\newblock ``Review of building energy modeling for control and operation''.
\newblock {\em Renewable and Sustainable Energy Review, {\bf 37}},
  p.~517–537.

\bibitem{Kim_Braun:2016}
Kim, D., Cai, J., Ariyur, K.~B., and Braun, J.~E., 2016.
\newblock ``System identification for building thermal systems under the
  presence of unmeasured disturbances in closed loop operation: Lumped
  disturbance modeling approach''.
\newblock {\em Building and Environment, {\bf 107}}, pp.~169 -- 180.

\bibitem{ZengIdentification_TCST:2019}
Zeng, T., and Barooah, P., 2020.
\newblock ``Identification of network dynamics and disturbance for a multi-zone
  building''.
\newblock {\em {IEEE} Transactions on Control Systems Technology, {\bf 28}},
  August, pp.~2061 -- 2068.

\bibitem{GuoAggregationEnB:2020}
Guo, Z., Coffman, A.~R., Munk, J., Im, P., Kuruganti, T., and Barooah, P.,
  2020.
\newblock ``Aggregation and data driven identification of building thermal
  dynamic model and unmeasured disturbance''.
\newblock {\em Energy and Buildings}, September.
\newblock in press, available online Sept 2020.

\bibitem{fritzson1998modelica}
Fritzson, P., and Engelson, V., 1998.
\newblock ``Modelica --- {A} unified object-oriented language for system
  modeling and simulation''.
\newblock In ECOOP'98 --- Object-Oriented Programming, E.~Jul, ed., Springer
  Berlin Heidelberg, pp.~67--90.

\bibitem{ASHRAE_handbook_applications:11}
{ASHRAE}, 2011.
\newblock The {ASHRAE} handbook : Applications ({SI Edition}).

\bibitem{liang2015mpc}
Liang, W., Quinte, R., Jia, X., and Sun, J.-Q., 2015.
\newblock ``{MPC} control for improving energy efficiency of a building air
  handler for multi-zone vavs''.
\newblock {\em Building and Environment, {\bf 92}}, pp.~256 -- 268.

\bibitem{taylor2007vav}
Taylor, S.~T., 2007.
\newblock ``{VAV} system static pressure setpoint reset''.
\newblock {\em ASHRAE journal, {\bf 6}}.

\bibitem{jorissen2018ideas}
Jorissen, F., Reynders, G., Baetens, R., Picard, D., Saelens, D., and Helsen,
  L., 2018.
\newblock ``{Implementation and Verification of the IDEAS Building Energy
  Simulation Library}''.
\newblock {\em Journal of Building Performance Simulation, {\bf 11}},
  pp.~669--688.

\bibitem{rouheiforpib:2001}
Roulet, C.-A., Heidt, F., Foradini, F., and Pibiri, M.-C., 2001.
\newblock ``Real heat recovery with air handling units''.
\newblock {\em Energy and Buildings, {\bf 33}}(5), pp.~495 -- 502.

\bibitem{ASHRAE_handbook_fund:17}
ASHRAE, 2017.
\newblock The {ASHRAE} handbook fundamentals ({SI Edition}).

\bibitem{SG_PB_Energy:11}
Goyal, S., and Barooah, P., 2012.
\newblock ``A method for model-reduction of non-linear thermal dynamics of
  multi-zone buildings''.
\newblock {\em Energy and Buildings, {\bf 47}}, April, pp.~332--340.

\bibitem{ASHRAE62-2016}
{ASHRAE}, 2016.
\newblock {ANSI/ASHRAE} standard 62.1-2016, ventilation for acceptable air
  quality.

\bibitem{SG_HI_PB_AE:2013}
Goyal, S., Ingley, H., and Barooah, P., 2013.
\newblock ``Occupancy-based zone climate control for energy efficient
  buildings: Complexity vs. performance''.
\newblock {\em Applied Energy, {\bf 106}}, June, pp.~209--221.

\bibitem{NSRDB}
National {S}olar {R}adiation {D}atabase ({NSRDB}).
\newblock \url{https://nsrdb.nrel.gov}.

\bibitem{Andersson2019}
Andersson, J. A.~E., Gillis, J., Horn, G., Rawlings, J.~B., and Diehl, M.,
  2019.
\newblock ``Casadi: a software framework for nonlinear optimization and optimal
  control''.
\newblock {\em Mathematical Programming Computation, {\bf 11}}(1), Mar,
  pp.~1--36.

\bibitem{wacbie:2006}
W{\"a}chter, A., and Biegler, L.~T., 2006.
\newblock ``On the implementation of an interior-point filter line-search
  algorithm for large-scale nonlinear programming''.
\newblock {\em Mathematical Programming, {\bf 106}}(1), Mar, pp.~25--57.

\bibitem{WilliamsWhy:2013}
Williams, J., 2013.
\newblock Why is the supply air temperature 55{F}?
\newblock
  \url{http://8760engineeringblog.blogspot.com/2013/02/why-is-supply-air-temperature-55f.html}.
\newblock Last accessed: Aug, 03, 2020.

\end{thebibliography}

%%%%%%%%%%%%%%%%%%%%%%%%%%%%%%%%%%%%%%%%%%%%%%%%%%%%%%%%%%%%%%%%%%%%%%%
%\appendix       %%% starting appendix
%\section*{Appendix A: Head of First Appendix}
%Avoid Appendices if possible.
%
%%%%%%%%%%%%%%%%%%%%%%%%%%%%%%%%%%%%%%%%%%%%%%%%%%%%%%%%%%%%%%%%%%%%%%%
%\section*{Appendix B: Head of Second Appendix}
%\subsection*{Subsection head in appendix}
%The equation counter is not reset in an appendix and the numbers will
%follow one continual sequence from the beginning of the article to the very end as shown in the following example.
%\begin{equation}
%a = b + c.
%\end{equation}

\end{document}